\documentclass[10pt,journal,compsoc]{IEEEtran}
\usepackage{amssymb}
\usepackage{amsmath}
\usepackage{amsthm}
\usepackage{array}
\usepackage{blindtext}
\usepackage{balance}
\usepackage{breakcites}
\usepackage{booktabs}
\usepackage{color}
\usepackage{comment}
\usepackage{caption} %\usepackage[font={small}]{caption}
\usepackage{epsfig,endnotes}
\usepackage{fancyvrb}
\usepackage{float}
\usepackage{fixltx2e}
\usepackage{graphicx}
\usepackage[hyphens]{url}
\usepackage{hyperref}
\usepackage{lipsum}
\usepackage{listings}
\usepackage{multirow}
\usepackage{mathtools,cuted}
\usepackage[normalem]{ulem}
\usepackage[numbers,super]{}
\usepackage{pifont} %used for tickmark and crosses.
\usepackage{paralist}
\usepackage{subfig}
\usepackage{url}
\usepackage{xspace}
\usepackage[table,xcdraw]{xcolor}
\usepackage[vlined,ruled]{algorithm2e}  %linesnumbered,
\newcolumntype{P}[1]{>{\raggedright\arraybackslash}p{#1}}
\newcommand{\nosemic}{\renewcommand{\@endalgocfline}{\relax}}% Drop semi-colon ;
\newcommand{\dosemic}{\renewcommand{\@endalgocfline}{\algocf@endline}}% Reinstate semi-colon ;

\newcommand{\ie}{\hbox{\em i.e.} }

\newcommand{\etal}{\hbox{\em et al.} }

\newcommand{\chame}{\textsc{Chameleon}\xspace}
\newcommand{\chames}{\textsc{Chameleon}'s\xspace}
\usepackage{romannum}

\colorlet{shadecolor}{gray!10}
\usepackage{filecontents}

\begin{document}

% \begin{frontmatter}
\title{A Praise for Defensive Programming: Leveraging Uncertainty for Effective Malware Mitigation}

\author{
\IEEEauthorblockN{Ruimin Sun\IEEEauthorrefmark{1}, 
Marcus Botacin\IEEEauthorrefmark{5},
Nikolaos Sapountzis\IEEEauthorrefmark{1},
Xiaoyong Yuan\IEEEauthorrefmark{1},   
Matt Bishop\IEEEauthorrefmark{3}, Donald E. Porter\IEEEauthorrefmark{4},\\ Xiaolin Li\IEEEauthorrefmark{2}, Andre Gregio\IEEEauthorrefmark{5} and Daniela Oliveira\IEEEauthorrefmark{1}}\\
\vspace{0.05in} 
\footnotesize{
\IEEEauthorblockA{University of Florida, USA \IEEEauthorrefmark{1} \{gracesrm,nsapountzis,chbrian\}@ufl.edu, \IEEEauthorrefmark{1}\{daniela\}@ece.ufl.edu}\\ 
\IEEEauthorblockA{University of North Carolina at Chapel Hill, USA \IEEEauthorrefmark{4}porter@cs.unc.edu}\\
\IEEEauthorblockA{University of California at Davis, USA \IEEEauthorrefmark{3}mabishop@ucdavis.edu}\\
\IEEEauthorblockA{Federal University of Parana, Brazil \IEEEauthorrefmark{5}\{mfbotacin,gregio\}@inf.ufpr.br}\\
\IEEEauthorblockA{AI Institute, Tongdun Technology, China \IEEEauthorrefmark{2}xiaolin.li@tongdun.net}
}
}

\maketitle

\pagestyle{plain}
\renewcommand{\thepage}{\arabic{page}}% Arabic page numbers

\begin{abstract}\label{sec:abstract}
A promising avenue for improving the effectiveness of behavioral-based malware detectors is to leverage two-phase detection mechanisms. 
% For example, there has been work combining fast but less accurate machine learning (ML) methods with high-accuracy but time-consuming deep learning (DL) models. 
% The main idea is when ML methods produce borderline classification for a piece of software, the software execution will be limited by an uncertain environment, and the DL methods will perform further analysis.
Existing problem in two-phase detection is that after the first phase produces borderline decision, suspicious behaviors are not well contained before the second phase completes.

% The main idea is to place software that receives borderline classifications from ML methods in an uncertain environment, and leverage DL models to do further analysis. The uncertainties will disproportionally affect poorly-written malware compared to well-written and resilient benign software, with the increasing demand for defensive programming.

This paper improves \chame, a framework to realize the uncertain environment. \chame offers two environments: standard---for software identified as benign by the first phase, and uncertain---for software received borderline classification from the first phase. The uncertain environment adds obstacles to software execution through random perturbations applied probabilistically. We introduce a dynamic perturbation threshold 
% which penalized known malicious behavior more intensively than innocuous behavior.
that can target malware disproportionately more than benign software. 
We analyzed the effects of the uncertain environment by manually studying 113 software and 100 malware, and found that 92\% malware and 10\% benign software disrupted during execution. The results were then  corroborated by an extended dataset (5,679 Linux malware samples) on a newer system. Finally, a careful inspection of the benign software crashes revealed some software bugs, highlighting \chame's potential as a practical complementary anti-malware solution.

\end{abstract}

% \end{frontmatter}

\begin{IEEEkeywords}
OS, Uncertainty, Malware, Fuzzing
\end{IEEEkeywords}

\section{Introduction}\label{sec:intro}

Real-time malware detection is challenging. The industry still relies on antivirus technology for threat detection \cite{kumar94,vigna98}, which is effective for malware with known signatures, but not sustainable for the massive amount of new malware samples released daily
(practical detection rates from 25\% to 50\%~\cite{bromium}). 
Thus, the AV industry started to rely on behavioral-based detectors built upon heuristics, which are more ``generic'' than signatures, but suffer from high false-positive rates~\cite{fireeye,paloalto}.
In a company, aggressive heuristics, i.e., those that are too focused \emph{on blocking suspicious software}, can interfere with employee's productivity, %everyday work-related activities of employees, 
resulting in employees overriding or circumventing security policies.

In addition, existing AV software mostly aims to identify the signature or monitor the runtime behavior through isolating the software for a while until a decision can be made \cite{fireeye, symantec13, mcafee09}. However, some software behavior may be hard to define. For example, malware may start by sleeping for five minutes or even longer and then perform malicious activities, or be only active during midnight and show benign behaviors for most of the time. These type of malware makes it hard for single step malware detector to give an accurate decision. 
 
Recently, two-phase hybrid detection methods are gaining attention \cite{li2015hybrid,sun2017learning} due to their capabilities in finding malware with complicated behaviors.
In \cite{sun2017learning}, the solution starts by using traditional machine learning models that are fast but not very accurate in its first-stage malware detection. If a borderline classification is received, modern deep learning methods that are accurate but time-consuming are performed for further analysis. However, the problem exists in this solution and any other two-phase detection methods is that, after the first phase gives a borderline decision, potential malware experiences no obstacles in executing its malicious behaviors before the second phase detector completes.

In this paper, we present \chame, a framework that separates the OS into two environments: standard and uncertain. The standard environment is a regular environment that all software starts execution from. In the uncertain environment, the software will experience probabilistic and random perturbations whose aim is to thwart the actions of potential malware while second phase analysis is under way. We provide a detailed description of the design and implementation of \chame, as well as new extensions to our framework.

The hypothesis is that the uncertain environment will mostly disturb malicious programs instead of benign ones. This is supported due to the increasing demand for defensive programming~\cite{sovtech} among software producing organizations.
Under this paradigm, poor-written malware code 
would be disproportionally more affected by the 
perturbations than well-written benign software, since defensive programming is a form of design intended to ensure the continuing operation of a piece of software under 
unforeseen circumstances, making the software behave in a reliable manner despite unexpected inputs or user actions.

In addition, malicious programs are not exquisite pieces of software overall---malware developers have to be able to quickly produce variants as AV signatures are created, causing them to be poorly coded. Malware also usually depend on specific configurations or installed applications to properly work, making them more prone to crashing due to uncertainties of the operating system (OS) it should run.

We evaluate the impact of uncertainty and unpredictability on actual malware samples compared to benign software
and discuss the reasons why the samples fail to address the unexpected
execution effects. \chame's strategy increases the cost of attacks, as writing malware using defensive programming requires
additional programming efforts and time.

In our evaluation of \chame~\cite{sun2017dose}, we manually inspected the execution and its effects of 100 samples of Linux malware and 113 common benign software from several categories. Our results show that at a perturbation threshold 10\% (i.e., a 10\% probability of perturbation for every system call that could be perturbed), intrusive perturbation strategies thwarted 62\% of malware, % to fail accomplishing their tasks, 
while non-intrusive strategies caused a failure rate of 68\%. At a perturbation threshold 50\%, the percentage of adversely affected malware increased to 81\% and 76\% respectively. With a 10\% perturbation threshold, the perturbations also cause various levels of disruption (crash or hampered execution) to approximately 30\% of the analyzed benign software. With a 50\% threshold, the percentage of software adversely affected raised to 50\%. We also found that I/O-bound software were three times more sensitive to perturbations than CPU-bound software.

Finally, we introduced an optional dynamic perturbation threshold to \chame. This threshold is computed so as to penalize more intensively software presenting known malicious behavior. Our analysis show that the application of such threshold caused 92\% of malware to fail and impacted only 10\% of benign software. Compared with a static threshold, this dynamic threshold improved in 20\% the number of benign software unaffected by the perturbations and caused 24\% more malware to crash or be hampered in the uncertain environment. We also analyzed the crash logs from benign software undergoing non-intrusive perturbations, and found that it was actually software bugs that caused the crashes.

\chame has the potential to advance systems security, as it can (i) make systems diverse by design because of the unpredictable execution in the uncertain environment, (ii) increase attackers' workload, and (iii) decrease the speed of attacks and their chances of success.
In this paper, we improved our work described in \chame \cite{sun2017dose}, and presented the following \textbf{new contributions:}

\begin{itemize}

\item We designed and implemented a dynamic perturbation threshold based on the behavior of software execution. We showed that such threshold is more effective than a static threshold in that it causes more adverse effects to malware execution and less impact to benign software.

\item We designed and implemented a fully automated Linux testbed\footnote{The testbed is publicly available at https://github.com/gracesrm/Chameleon-malware-testbed} for collecting system call traces (at kernel level) from malware and benign software when these software is under perturbations. Such testbed can be leveraged to analyze benign software behavior under OS misbehavior and help developers pinpoint portions of their software that are sensitive to misbehavior, thus leading to more resilient software. 

\item We further collected 5,679 Linux malware samples and analyzed this extended malware dataset on a new version system. The results corroborated our findings on previous small sample set, and indicated \chame's capability of standing the test of time.

\item We provided the results of analysis of case studies of applications running under \chame, including malware using three evasive stalling techniques, and commonly used benign software (e.g. vim, tar, Mozilla Firefox and Thunderbird) affected.
\end{itemize} 
 
This paper is organized as follows. Section \ref{sec:threat} describes our threat model and assumptions. Section \ref{sec:system} describes in detail \chame's design and implementation, including the newly proposed dynamic perturbation threshold. Section \ref{sec:eval} describes \chame's security and performance evaluation, including our analysis of causes of crashes for benign software in the uncertain environment. Section \ref{sec:discuss} discusses and summarizes \chame's results and limitations. Section \ref{sec:relatedwork} summarizes related work on malware detection, software diversity, and attempts on unpredictability as a security mechanism. Section \ref{sec:conclusion} concludes the paper.

\section{Threat Model and Assumptions}\label{sec:threat}
% \chames goal is to provide an environment that rate-limits the effects of potential malware, while more time-consuming deep analysis is underway. 
\chames protection is designed for corporate environments, which have adopted the practice of controlling software running at their perimeters \cite{paloalto}. 
% Organizations face the challenge of needing to enforce perimeter security, while causing minimum interference to employees' primary tasks. The combination of fast, preliminary classification by traditional machine learning methods and more time-consuming and accurate deep analysis for borderline cases can help address this challenge. 

%is to provide an environment that rate-limits the effects of potential malware while DL methods are in operation, which can be time-comsuming. Organizations demanding high security level and applying detectors using aggressive heuristics, such as \emph{erring on the side of blocking suspects}, can interfere with work-related activities of employees', resulting in their overriding or circumventing security policies \cite{kruegel10,canali12,fireeye}.

We assume that if an organization is a target of a well-motivated attacker, malware will eventually get in (e.g., spear-phishing). 
% A classic scenario is when a C-level personnel of a targeted organization falls victim to a spear-phishing email attack, thereby causing a backdoor to be installed in one of the computers of the victim's company. 
If the malware is zero-day, it will not be detected by any signature-based antivirus (AV). If the malware receives a borderline classification by behavioral-based detectors, it might lurk inside the organizations for extended periods of time.
% It also behaves in a way that does not raise red flags for a behavior-based detector. Further, a mis-configuration in the administrator's software restriction policies allows the software to run. 
% In a standard OS, this piece of malware would infiltrate and compromise the organization. 
With \chame, if this piece of malware might receive a borderline classification at some point by a conventional machine learning detector, it would then be placed in the uncertain environment. In this environment the malware would encounter obstacles and delays to operate, % works in a hampered fashion,
while more time and resource-consuming deep analysis is underway to definitely flag it as malicious.

% We assume that whitelisted software receiving a borderline classification by a conventional machine learning detector can be an indication of a malware. %We assume that all malware will be identified as suspicious by traditional ML detectors and benign software wrongly labeled (false-positives) will be identified by a deep-learning analysis in the uncertain environment.
%We also assume that if an organization is a target of a well-motivated attacker, malware will eventually get in. A classic scenario is a C-level personnel of a targeted organization who falls victim of a spear phishing email attack causing an APT backdoor to be installed in one of the computers of the victim's company. The malware is zero-day and is not detected by any antivirus. It also behaves in a way that does not raise red flags for a behavioral-based detector. In addition, a misconfiguration in the administrator software restriction policies allows this malware to run. In a standard OS, this APT would initiate a devastating attack on the organization. In the uncertain environment, the APT backdoor executes in a hampered fashion. It will run slower, will not have full connectivity to its command and control server (C\&C), and will experience transient unavailability to needed resources, such as system files. Thus, the uncertain environment will not only hampers the APT, but will also buy time for behavioral detectors to definitively classify the APT's actions as malicious.
% It is worth noting that \fixme{"of note" to start a sentence tends to be informal} 
\chame does not compete with standard lines of defenses, such as conventional AVs, behavioral-based detectors, and firewalls, but actually equips these solutions with a safety net in case of misdiagnosis.

\section{Design and Implementation}\label{sec:system}
We designed and implemented \chame for the Linux OS. %Uncertainty was implemented as a set of unpredictable perturbations at the system call layer. 
\chame offers two environments to its processes: (i) a standard environment, which works predictably as any OS, and (ii) an uncertain environment, where a subset of the OS system calls for selected processes undergo unpredictable perturbations. %The main idea for \chame is to place mission critical and approved software in the standard environment, and everything else (may include stealthy malware) in the uncertain environment. 

The key insight is that perturbation in the uncertain environment will hamper malware's chances of success,
as some system calls might return errors in accessing system resources (e.g. network connections or files) or cause malware execution delays. 

\subsection{The Perturbation Set}\label{sec:perturbation}
Our first step was deciding which system calls were good candidates for perturbation. We relied on Tsai et al.'s study~\cite{tsai16support}, which ranked Linux system calls by their likelihood of use by applications. Based on these insights, we selected 37 system calls for the perturbation set to represent various OS functionalities relevant for malware (file, network, and process-related). Most of these system calls (summarized in Table \ref{tab:syscallset}) are I/O-bound, since I/O is essential to nearly all malware, regardless of its sophistication level.  

We introduced new versions for all system calls in the perturbation set. %Because the Linux system call table is read-only in more recent kernel versions, we used the Linux System.map file to discover the address of the system call table, and changed its memory page to read-write mode. 
For each system call {\tt orig\_<syscall\_name>} in the perturbation set, \chame altered the corresponding table entry to point to {\tt my\_<syscall\_name>}, in which perturbations were added if the software was executing under the uncertain environment.

\begin{table}[!ht]
\centering
\scriptsize
\caption{\footnotesize{System call perturbation set.}}
\label{tab:syscallset}
\vspace*{-\baselineskip}
\begin{tabular}{|l|l|}
\hline
\textbf{Category}  & \textbf{System call} \\ \hline
\begin{tabular}[c]{@{}l@{}}File- \\ related\end{tabular}    & 
\begin{tabular}[c]{@{}l@{}}sys\_open, sys\_openat, sys\_creat, sys\_read, \\
sys\_readv, sys\_write, sys\_writev,  sys\_lseek, \\
sys\_close, sys\_stat, sys\_lstat, sys\_fstat, \\
sys\_stat64, sys\_lstat64, sys\_fstat64, sys\_dup, \\ 
sys\_dup2, sys\_dup3, sys\_unlink, sys\_rename\end{tabular} \\ \hline
\begin{tabular}[c]{@{}l@{}}Network-\\ related\end{tabular} & \begin{tabular}[c]{@{}l@{}}sys\_bind, sys\_listen, sys\_connect, sys\_accept, \\ sys\_accept4, sys\_sendto, sys\_recvfrom, \\ sys\_sendmsg, sys\_recvmsg, sys\_socketcall\end{tabular}   \\ \hline
\begin{tabular}[c]{@{}l@{}}Process- \\ related\end{tabular} & 
\begin{tabular}[c]{@{}l@{}}sys\_preadv, sys\_pread64,\\ sys\_pwritev, sys\_pwrite64, \\
sys\_fork, sys\_clone, sys\_nanosleep \end{tabular}   \\ \hline
\end{tabular}
\end{table}
 
\subsection{Perturbation Strategies}

We introduced two sets of perturbation strategies. The first, \textit{non-intrusive}, perturbed software execution within the OS specification.
The second, \textit{intrusive}, could cause corruption in the software execution.
From the end user point of view, intrusive strategies might cause functionalities to be temporarily unavailable. Non-intrusive strategies might cause software to run slower.

\subsubsection{Non-intrusive Perturbation Strategies}

\noindent 
\textit{1. System call silencing with error return}: The system call immediately returns an error value randomly selected from the range [-255, -1]. This perturbation strategy can create difficulties for the execution of the process which do not handle errors well. Further, this strategy can cause transient unavailability of resources, such as files and network, creating difficulties for certain types of malware to operate. In this perturbation type, all error returns are within the OS specification, i.e., the expected set of return values.

\noindent
\textit{2. Process delay}: The system call injects a random delay during its execution to delay potential malware execution. It can create difficulties in timely malware communication with a C\&C for files ex-filtration, as well as prevent flooders from sending enough packets in a very short time, rate-limiting DoS attacks. The delay range was chosen as a random number within [0,0.1] as an experimental range. 
A delay longer than 0.1s could cause network applications to timeout and terminate early.

\noindent
\textit{3. Process priority decrease}: The system call decreases the dynamic process priority to the lowest possible value, delaying process scheduling. 

\subsubsection{Intrusive Perturbation Strategies} 

\noindent
\textit{1. System call silencing}: The system call immediately returns a value (without being executed) indicating a successful execution.

\noindent
\textit{2. Buffer bytes change}: The system call experiences an increase or decrease in the size of a buffer passed as parameter. It can be applied to all system calls with a buffer parameter, such as {\tt sys\_read}, {\tt sys\_write}, {\tt sys\_sendto} and {\tt sys\_recvfrom}. This strategy can corrupt the execution of malicious scripts, thus making exfiltration of sensitive data more difficult. This strategy also targets viruses, which can be adversely affected by the disruption of the buffer with a malicious payload trying to be injected into a victim's ELF header---the victim %\fixmedp{huh?  Do you mean the jump tables, or just teh buffer itself?  or does the buffer include an elf binary?} 
process may get corrupted and lose its ability to infect other files.

\noindent
\textit{3. Connection restriction}: The strategy changes the IP address in {\tt sys\_bind}, or limits the queue length for established sockets waiting to be accepted in {\tt sys\_listen}. The IP address can be randomly changed, which will likely cause an error, or it can be set to the IP address of a honeypot, allowing backdoors to be traced.

\noindent
\textit{4. File offset change}: The strategy changes a file pointer in the {\tt sys\_lseek} system call so that subsequent invocations of {\tt sys\_write} and {\tt sys\_read} will access unpredictable file contents within a specified, configurable range.

\subsection{System Architecture}\label{sec:arch}

\begin{figure*}[!ht]
\centering
\includegraphics[width=\columnwidth]{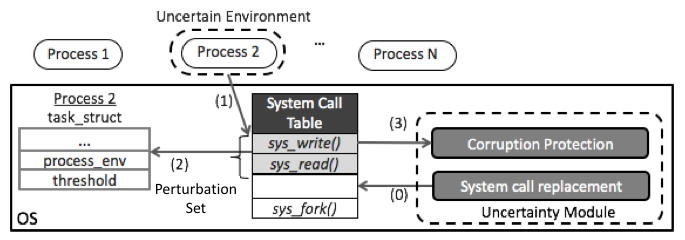}
\caption{
    \footnotesize{
    \chame's architecture. When a process running in the uncertain environment invokes a system call in the perturbation set (Step 1), the \textit{Uncertainty Module} checks if the process is running in the uncertain environment (Step 2), and depending on the execution of the corruption protection mechanism (Step 3), randomly selects a perturbation strategy to apply to the system call. The corruption protection mechanism prevents perturbations during accesses to critical files, such as libraries.}
    }
\label{fig:arch}
\end{figure*}

The key component of \chame is a kernel-level uncertainty module (see Figure \ref{fig:arch}) which is responsible for implementing the uncertain environment. Specifically, this module (i) hooks into the Linux system call table \footnote{For system call monitoring, we choose to hook into the system call table through a loadable kernel module rather than using {\tt strace}, because malware with anti-analysis techniques may stop executing when {\tt strace} is detected.} to replace system calls in the perturbation set with new versions of system calls that apply perturbations (Step 0 in Figure \ref{fig:arch}), (ii) monitors system calls invoked by processes in the uncertain environment, and (iii) applies perturbations to the system calls when required. The perturbation strategies are chosen randomly, and applied probabilistically, in the uncertain environment.

\chame added the following fields to the Linux {\tt task\_struct}:

{\tt process\_env}: a flag informing whether or not the process should run in the uncertain environment. 

{\tt fd\_list}: a list of critical file descriptors during process execution. Applying perturbation to system files, such as library or devices, will likely cause the process to crash. Thus, \chame does not apply perturbations to system calls manipulating those file descriptors (see Section~\ref{sec:corrupt} for more details).

{\tt strategy\_set}: a flag informing the type of perturbation strategies the process should undergo: non-intrusive or intrusive.

{\tt threshold}: an integer informing 
the probability that a system call from the perturbation set invoked by a process in the uncertain environment will undergo perturbation. In other words, the threshold represents the strength of the perturbation to be applied to a system call. The higher the threshold, the higher the probability that a perturbation strategy will be applied. By default, threshold 10\% is used.

\chame's operation is illustrated in Figure \ref{fig:arch}. Consider \texttt{Process 2}, running in the uncertain environment invoking {\tt sys\_write} (Step 1), which belongs to the perturbation set. Thus, as determined by the perturbation threshold, a perturbation strategy can be applied to its execution. During \chame's operation, the hooked 
system call first inspects \texttt{Process 2}'s environment 
and finds that it runs in the uncertain environment (Step 
2). Next, {\tt sys\_write} runs the corruption protection 
mechanism (see Section~\ref{sec:corrupt} for details) to make sure that 
no perturbation will occur if the system call is accessing a 
critical file (Step 3). If {\tt sys\_write} is not accessing 
a critical file, \chame 
decides based on the probabilistic threshold whether or not a perturbation 
should be applied. If a perturbation is to be applied, {\tt 
sys\_write} randomly selects one of the perturbation strategies that can 
be applied to its execution.

\subsection{Dynamic Perturbation Threshold} \label{sec:threshold}
As explained earlier, the perturbation threshold, denoted as $T_{syscall}$ hereafter, represents the \textit{probability} that a system call belonging to the Perturbation Set and invoked by a process in the uncertain environment will undergo perturbations. 
In our original work~\cite{sun2017dose}, we assumed the same default perturbation threshold for all system call invocations and all $syscall$ types when applicable. 
In this section, we go a step beyond, and propose threshold $T_{syscall}$ that will change based on the execution context and in a per-system call, per-process fashion: it will be higher for processes exhibiting known likely malicious behavior (to ensure more system calls invoked by a malicious software are perturbed) and will be lower for likely benign processes. 
To that end, we introduce a dynamic \textit{perturbation threshold} that changes based on the system call type and invocation context of a given process. The goal is to reduce the chance for benign software to be false positively killed.

In the remainder of the section, we describe two families of behaviors that are considered suspicious in the literature \cite{elfinjection,dupcall,wen2017cyberspace,garfinkel2003traps}.  
First, in Section~\ref{sec:sig-threshold} we consider the \textit{``signature"} of the system call, \textbf{\textit{namely a type of system call along with its parameters}} (Behavior family A). The literature reports some signatures for a few malicious invocations of system calls \cite{elfinjection,dupcall,wen2017cyberspace,garfinkel2003traps}. 
% ~\cite{aafer2013droidapiminer,canali12,kirda2006behavior}. 
To that end, if a system call is going to be invoked with a combination of parameters matching the signature of a known likely malicious behavior, \chame increases the threshold for that particular system call invocation to some pre-defined threshold, thus affecting the system call execution on-the-fly. Second, in Section~\ref{sec:freq-threshold} we consider the invocation \textit{frequency} of the system call (Behavior family B). Specifically, if a system call is invoked in an abnormally frequent manner and matches a likely malicious behavior (e.g. DoS), \chame perturbs its invocation with a probability that changes according to the invocation frequency. 
For each family of behaviors (signature or frequency), we identify examples of some malware representative of the behavior, but one can extend these families to consider more malicious behaviors. 
In Table \ref{tab:var} we summarized the notation used. %The input parameters are labeled with *.

%As an extension of our original work on \chame \cite{sun2017dose}, representing a new contribution of this paper, we introduce a per system call perturbation threshold that dynamically changes based on the execution context. As explained previously, this threshold represents the \textit{probability that a system call invoked by a process in the uncertain environment will undergo perturbation}, and \chame applies the same default threshold value $t_\text{d}$ to all system calls without considering execution context. 

%However, in practice, based on the execution context, system calls can exhibit benign or suspicious \cite{aafer2013droidapiminer,canali12,kirda2006behavior} behaviors, and this calls for a threshold that can be dynamically adapted to such behaviors. In the remainder of the section, we summarize two families of behaviors that are highly suspicious, and explain how we derive the thresholds based on a per-system call signature and a per-system call frequency for these two families.

\subsubsection{Behavior family A: a signature-based threshold} \label{sec:sig-threshold}
In the first family of malicious activities, we propose three system call \textit{signatures} for three different malicious behaviors. 
%The first family of suspicious behaviors can be represented by the \textit{signature} of a certain system call and its parameters. 
We found that these \textit{signatures} are prevalent in approximately 95\% of the malware samples, and occurred in only 5\% of the benign software samples in the system call traces we collected. 
Therefore, it is plausible to assume that a process invoking such behavior is likely malicious.

\vspace{0.1cm}
\noindent \textit{(A1. ELF Header Injection)}: A strategy often employed
by malware samples to get privileged access to a system is to replace
existing binary contents with malicious payloads, thus benefiting
from root and/or whitelist execution permissions previously attributed
to the affected binary. This kind of injection can be correlated to the
signature {\tt sys\_write("$\backslash$177ELF")}~\cite{elfinjection},  
which considers the ELF (Executable and Linkable Format) magic number 
as argument for the {\tt sys\_write} system call.

\vspace{0.1cm}
\noindent \textit{(A2. I/O Redirection)}: Malware samples often rely 
on redirecting standard I/O descriptors to implement their malicious
behaviors. The keyboard descriptor is often redirected by keyloggers 
to allow data collection. The screen output descriptor is often redirected 
by remote shells to hide attacker's commands. I/O redirection can be
correlated to signatures targeting system calls that modify the standard I/O 
descriptors, such as {\tt sys\_dup(fd)}, where $fd=\{0,1,2\}$ is used as 
a stub to obtain the victim server's standard input, output and error,
respectively~\cite{dupcall}. 

\vspace{0.1cm}
\noindent \textit{(A3. Replacing system binaries)}: Rootkits
often replace (rewrite, remove or unlink) existing legitimate 
applications by malicious/trojanized versions so that they can 
violate some security policy or hide malware 
traces~\cite{garfinkel2003traps}. In this sense, system binaries 
are the most targeted files by rootkits~\cite{wen2017cyberspace},
given their ability to provide users with system information, 
such as the running processes list. This behavior can be correlated to the system call signatures {\tt sys\_unlink("path")} or {\tt sys\_rename("path")}, and the "path" refers to the location of system binaries, e.g. "/bin", and "/usr/bin".

\vspace{0.1cm}
For the sake of generality, we introduce three different thresholds for each cases of behavior family A, as they might refer to different levels of likelihood of malicious activity. We assume these thresholds to be input parameters, that can be changed based on a variety of factors, such as the running application, the requirements of an organization, etc.
Specifically, we assume these thresholds to be $t_{A1}, t_{A2}, t_{A3}$, respectively, where 
$A1=sys\_write("\backslash177ELF")$, $A2= sys\_dup(fd)$, $A3=\{sys\_unlink("path") \textbf{ or }sys\_rename("path")\}$.

Note that \chame does not kill the corresponding software nor it increases the perturbation threshold to 100\% ($t_{Ai} < 1$) because benign software might also exhibit such behaviors with a small probability, and \chame expects benign software written with good quality to be resilient to those perturbations.

\begin{table}[!ht]
\centering
\scriptsize
\caption{\footnotesize{Notation. These are per-process parameters.}}
\label{tab:var}
\vspace{-\baselineskip}
\begin{tabular}{|c|l|}
\hline
\textbf{(Subscripts)}  & \\ \hline
${A1}$ & System call Behavior A1: \textit{sys\_write("/177ELF")} \\ \hline
${A2}$ & System call Behavior A2: \textit{sys\_dup(0)}, \textit{sys\_dup(1)}, \textit{sys\_dup(2)}\\ \hline
${A3}$ & System call Behavior A3: \textit{sys\_unlink("path")}, \textit{ sys\_rename("path")}\\ \hline
%${Ai}$ & Subscript for the signature of Behavior family A i.e., A1, A2 or A3 \\ \hline
$syscall$ & System call type $syscall$ \\ \hline
$\text{max,min}$ & The maximum/minimum allowed value of an input \\ \hline

\textbf{Inputs}   & \\ \hline

$t_{Ai}$ & Fixed perturbation threshold for Behavior family $A$ \\ \hline
$p$ & Coefficient of the relationship between $F_{syscall}$ and $T_{syscall}$\\ \hline
${N}_\text{min}$ & Minimum allowed ${N} $ for Behavior B \\ \hline
$F_\text{min}$ & Minimum allowed $F_{syscall}$ for Behavior B \\ \hline
$t_\text{max}$ & Maximum allowed value for all perturbation thresholds\\ \hline

\textbf{Variables}  & \\ \hline
$T_{syscall}$ & Perturbation threshold for the system call type $syscall$ \\ \hline
$ n_{syscall} $ &  Total number of system calls invoked for type $syscall$\\ \hline
$F_{syscall}$ &  Invocation frequency for the system call type $syscall$ \\ \hline
${N} $ & Total number of system calls invoked \\ \hline
\end{tabular}
\end{table}

\subsubsection{Behavior family B: a frequency-based threshold} \label{sec:freq-threshold}
The second family of suspicious behaviors can be represented by the frequent invocation of a certain type of system call during a process execution, as shown by Ptacek and Newsham~\cite{ptaceknewsham}. 
The goal of this family of malicious behaviors is to aggressively consume system resources until the system cannot function for some or all of its legitimate requests (e.g. DoS). In the following, we list two suspicious behaviors belonging to this family and the corresponding system calls that are frequently invoked. % (more behaviors can be added per user configuration).

\vspace{0.1cm}
\noindent \textit{(B1. Network flooding.)} This can represent malware seeking to make network resources unavailable to its intended users by temporarily or indefinitely disrupting services of a host connected to the Internet. Network flooding is typically accomplished by flooding the targeted machine with frequent invocations of {\tt sys\_sendto} or {\tt sys\_recvfrom}.

\vspace{0.1cm}
\noindent \textit{(B2. Fork bomb.)} This can represent malware forking processes infinitely until the system runs out of memory. 
Once the fork bomb is activated, it may not be possible to resume normal operation without rebooting the system. A fork bomb, as the name implies, is characterized by frequent invocations of {\tt sys\_fork}. 

In the remainder of the section, we: 
\textit{first}, explain when a system call is considered as Behavior family B based on the invocation frequency, and \textit{then}, we derive a formula that dynamically adapts their threshold based on the system call invocation frequency. 
Let us denote, in a process, the number of times that the system call type \textit{system call} is invoked as $n_{syscall}$ and the current total number of system calls invoked as ${N}$. 
% In terms of \textit{system call} is a system call sequence, $n_{system call}$ aggregates the times of invocations for every system call in the sequence.
Then, we introduce the {invocation frequency} of the system call \textit{syscall} as\footnote{Note that the invocation frequencies of all system calls should sum to $1$, i.e. $\sum_{system call} F_{syscall} = 1$.}:
%For \textit{Behavior 1}, given that different programs invoke different number of system calls, the first step is to define the frequency of the invocation. We derive the invocation frequency, namely $0 \leq R \leq 1$, as follows:
\begin{align}
F_{syscall} = \frac{n_{syscall}}{{N}}. 
\label{eq:r}
\end{align}

We assume that a system call $syscall$ falls into Behavior family B when both of the following conditions occur (i) the invocation frequency $F_{syscall}$ is higher than a pre-defined minimum frequency $F_\text{min}$ (namely $F_{syscall} \geq F_\text{min}$), and 
(ii) the total number of system calls invoked $N$ exceeds a minimum number $N_\text{min}$ (namely $N \geq N_\text{min}$), where $F_\text{min}$ and $N_\text{min}$ are both input parameters. The first condition ensures that the system call invocation frequency is high enough to be considered as suspicious. %, e.g., one invocation of {\tt sys\_bind} out of a total of 1,000 system call invocations (1/1000 = 0.1\%) is not considered as Behavior family B. 
The second condition ensures that the software has launched itself and the first system calls invoked by a process are not mistakenly perturbed. %, e.g. a total of five system call executions {{\tt sys\_execve}, {\tt sys\_open}, {\tt sys\_open}, {\tt sys\_open}, {\tt sys\_open}} with $F_{sys\_{open}} = 4/5 = 0.8$ is not considered as Behavior family B. 

We now derive the per-system call threshold $T_{syscall}$ as a function of frequency. The goal is to increase the threshold $T_{syscall}$ as the frequency $F_{syscall}$ increases, because the higher $F_{syscall}$ the more likely the process is exhibiting one of the behaviors in Behavior family B. One could use different functions for that objective, and ``penalize" the threshold $T_{syscall}$ (and further the probability of a system call being perturbed) on $F_{syscall}$ in a linear, quadratic, cubic, etc. manner. % (the higher the power the higher the penalty increase). 
Different powers correspond to different penalty increases. 
Without loss of generality, we chose to use linear function with coefficient $p$ as an input parameter, and the threshold of \textit{system call} is defined as follows: 
\begin{align}
T_{syscall}=\begin{cases}
  %  t_d, & \text{if $R \leq R_{min}$ or $N_{system call} \leq N_{min}$},\\ 
    p \cdot F_{syscall}, & \text{if $p\cdot F_{syscall} \leq t_\text{max}$ },\\ %and  $R \geq R_{min}
    t_\text{max}, & \text{otherwise}.
  \end{cases} 
\label{eq:T}
\end{align}

We want to keep $T_{syscall} \neq 1$ to prevent the scenario where all system calls are perturbed and software crashes at the very beginning, and thus $T_{syscall}$ takes values $0 \leq T_{syscall} < 1$. 
The input parameter $0 \leq t_\text{max} < 1$ dictates the maximum value that $T_{syscall}$ can obtain. 
Specifically, the threshold $T_{syscall}$ increases linearly on the product between the coefficient and the invocation frequency, namely $p \cdot F_{syscall}$, when the latter is less than $t_\text{max}$ (see first branch of Eq.~\ref{eq:T}). 
On the other hand, if the product exceeds the maximum value $t_\text{max}$, then the threshold is $t_\text{max}$ (see second branch of Eq.~\ref{eq:T}). $t_\text{max}$ allows the user to customize the maximum threshold, namely the strongest perturbations desired for the software. Note that, as we need to keep $0 \leq T_{syscall} < 1$, $p$ can take values $0 \leq p < \frac{1}{F_{syscall}}$.

We show here an example for the second family of suspicious behaviors, with $F_\text{min}=0.7$. Assuming a flooding attack on a machine, that invokes a {\tt sys\_recvfrom} system call with invocation frequency $F_\text{syscall}=0.9$, and has an input coefficient parameter $p=0.8$. Then, the threshold will be $T_\text{syscall}=0.8*0.9=0.72$. 
This is classified as the second family of suspicious behaviors and the {\tt sys\_recvfrom} system call has 72\% probability to encounter perturbations. 
It is clear that, the higher the frequency $F_{syscall}$ or the parameter $p$, the higher the $T_{syscall}$ and thus the more frequent the software will encounter perturbations. 

Note that a certain system call can simultaneously belong to the two families of suspicious Behavior families A and B.
For example, system call {\tt sys\_write("$\backslash$177ELF")} that is hallmark for the Behavior family A may also have high invocation frequency ($F_\text{sys\_write} > F_\text{min}$) suggesting that it also belongs to Behavior family B. 
In this case, each behavior would suggest a threshold: the first family would suggest $ t_{A1}$ and the second family would suggest $T_\text{sys\_write}$. 
As the probability for benign software to exhibit two suspicious behaviors at the same time is relatively small, we chose to use the higher threshold between the two behaviors i.e. $T = \max  \{ {T_{sys\_write},t_{A1}}$\} for the software under consideration.

For each of the input parameters, we chose an experience value to carry out our experiments. Without loss of generality, we chose $t_{A1}= t_{A2}= t_{A3}= t_\text{max}$ for simplicity of the experiment. We chose $t_\text{max} = 0.95$ because the strongest perturbations should still allow software to run and exhibit its behaviors, namely $t_\text{max}$ should be close to but smaller than 1. We chose $N_\text{min} = 100$ because most of the software can finish loading its libraries by the time of 100 system calls are invoked. We chose $F_\text{min} = 0.7$ because the least aggressive DoS software can still invoke a certain type of system call as frequent as to occupy 70\% of all the system calls. We chose $p=1.2$ (since $F_\text{min} = 0.7$ the maximum value of $p$ is $1.42$ - see Eq.~\ref{eq:T}). 
These values cause less perturbations in benign software while affected malware more, as our evaluation results showed in Section \ref{sec:eval}. For system calls which do not exhibit behaviors, as described by behavior families A and B, the default threshold will be used.

\subsection{Corruption Protection Mechanism} \label{sec:corrupt}
The uncertainty module employs a corruption protection mechanism to prevent perturbations while a process in the uncertain environment is accessing critical system files, which might cause process to crash at a very early stage. The files are identified through file descriptors, created by {\tt sys\_open}, {\tt sys\_openat} and {\tt sys\_creat}, and are deleted by {\tt sys\_close}. System calls whose parameters are file descriptors, such as {\tt sys\_lseek}, {\tt sys\_read} and {\tt sys\_write}, are under this protection mechanism. %It accesses the \emph{fd\_list} data structure, which maps processes' PIDs to file descriptors. 
These protected files are determined by an administrator and 
tracked by setting an extended attribute
in the file's inode in the {\tt .security} namespace (a similar strategy
is employed by SELinux~\cite{linuxse}).

When a process running in the uncertain environment opens a file with a path name beginning with the name or containing keywords of the critical directories, the file descriptor ({\tt fd}) %\fixmedp{Consider using tt or em rather than math mode, as each char is treated by latex as a separate variable, which looks a little weird in the pdf} 
is added to a new per-process data structure $fd\_list$. Later, when this process invokes {\tt sys\_read} or {\tt sys\_write} referring to an $fd$ in $fd\_list$, the protection mechanism will prevent perturbation strategies from being applied to these system calls.

Algorithm \ref{alg:strategywrite} shows how the OS applies the perturbation strategies on {\tt sys\_write}. First, the following conditions are checked: (i) the process is running in the standard environment ($process\_env == 0$), and (ii) the targeted file descriptor is a critical system file. If either of the two conditions is true the system call runs normally. Otherwise, the system call updates its execution counters of the current process (\ie the total number of system calls invoked ${N}$ and the total number of {\tt sys\_write} invoked ${n}_\text{sys\_write}$) and check whether {\tt sys\_write} exhibits frequent invocation based on the input parameter $F_{min}$. 
Then the algorithm generates a random number in the range [0,1], and if the number is smaller than the threshold, the system call undergoes perturbation. 

The algorithm will randomly select one of the perturbation strategies based on the strategy type. If non-intrusive strategies are selected, one of the following strategies will be randomly selected for execution: \emph{System call silencing with error return}, \emph{Process delay}, or \emph{Process priority decrease}. If {\tt sys\_write} is silenced, a random error code is returned, so that the process knows that an error occurred. If \emph{Process delay} is chosen, the algorithm randomly selects a delay for the system call execution in the range $[0, 0.1s]$. Our experiment show that a delay longer than 0.1 second will cause the program to timeout and the software to terminate at an early stage. If \emph{Process priority decrease} is selected, the algorithm decreases the process priority to the minimum. If intrusive strategies are selected, perturbation strategies \emph{System call silencing}, or \emph{Buffer bytes change} will be randomly selected. If \emph{System call silencing} is selected, \chame {\tt sys\_write} will return the buffer size without actually writing to the file. Otherwise, \chame will change the buffer bytes and return the new length of the buffer.

\begin{algorithm}[!ht]
\footnotesize
\SetAlgoLined
\caption{\footnotesize Applying perturbations to {\tt sys\_write()}}
\SetKwProg{Fn}{Function}{}{}
\Fn{long \textbf{my\_sys\_write}(fd, buf, size)}{
	\eIf{process\_env == 0 \textbf{or} corruption\_protection(sys\_write, pid, fd\_list)}{
		\Return orig\_sys\_write(fd, buf, size)\;	
	}{
    	boolean top\;
        freq = isFrequentCalls($N$++, $n_{sys\_write}$++)\;
		updateThreshold(freq)\;
		\If{$(random(0.0, 1.0) > threshold)$}{
			\Return orig\_sys\_write(fd, buf, size)\;
		}
		\If{$strategy\_set == Non-intrusive$}{
		    $strategy$ = random(1,3)\;
            \uIf{$strategy = 1$}{ \tcc{System call silencing with error return}
                \Return random(-255, -1)\;
            }
            \uElseIf{$strategy =2$}{ 	\tcc{Process delay}
                delay(random(0, MAX\_DELAY))\;
                return orig\_sys\_write(fd, buf, size)\;
            }
            \Else{ \tcc{Process priority decrease}
                decrease\_current\_priority()\;
                return orig\_sys\_write(fd, buf, size)\;
            }	
		}\Else{ \tcc{$strategy\_set == Non-intrusive$}
		    $strategy$ = random(1,2)\;
        	\uIf{$strategy = 1$}{ \tcc{System call silencing}
                \Return size\;
            }
            \Else{ \tcc{Buffer bytes change}
            	newbuf = RandomBytes(buf)\;
                return orig\_sys\_write(fd, newbuf, size)\;
            }
        }
	}
}
\setcounter{AlgoLine}{0}
\label{alg:strategywrite}
\end{algorithm}

\section{Evaluation}\label{sec:eval}
The goal of our evaluation was to discover the impact of \chame's uncertain environment in affecting malware and benign software behavior. We considered security, performance, and software behavior to answer the following research questions: (i) how will the uncertain environment with perturbation strategies affect software execution? (ii) is the per-process, per-system call perturbation threshold more effective than a static threshold? (iii) how different strategies impact malware in the uncertain environment?, and (iv) how can benign software be more resilient in the uncertain environment?

Our evaluation leveraged a collection of 113 software including common software from GNU projects \cite{gnuproject}, SPEC CPU2006 \cite{cpu2006} and Phoronix-test-suite \cite{pts} (47 I/O-bound and 66 CPU-bound). The malware samples used in our evaluations were selected from THC \cite{thc} and VirusShare \cite{virusshare}.

Our selection criteria was to have a diverse software dataset, which at the same time, could allow timely \textit{manual} analysis of all aspects of execution. For certain cases, the outcome of software execution can only be analyzed via manual inspection (e.g., the outcome of files produced by a text editor under the uncertain environment). Other reasons for the need to perform \textit{manual} analysis were as follows. First, we needed to reverse engineering malware binaries to discover whether the malware sample was a self-contained \textit{desktop} binary, which libraries and versions were required for the malware execution, which input parameters were required for execution, and to correctly install the specific software and version that a certain malware would inject payload into. Second, we needed to prepare the malware with all its required resources, such as installing libraries, setting up the environment, and launching the victim software.

We analyzed 100 malware samples belonging to different categories (22 flooders, 14 worms, 15 spyware, 24 Trojans, and 25 viruses). The samples contained executables built on both x86 and x86\_64 systems. All the malware and benign software used in our experiments are detailed in Table 1 and Table 2 in the Appendix \cite{appendix}. 

We deployed and evaluated \chame on two virtual machines (VMs) both running Ubuntu 12.04 with 1GB RAM, 30GB Hard Disk, and one processor. One VM used x86 architecture, and the other used x86\_64 architecture.

\subsection{Testbed and Data Collection}
\begin{figure*}[htb]
	\centering
	\includegraphics[width=0.7\textwidth]{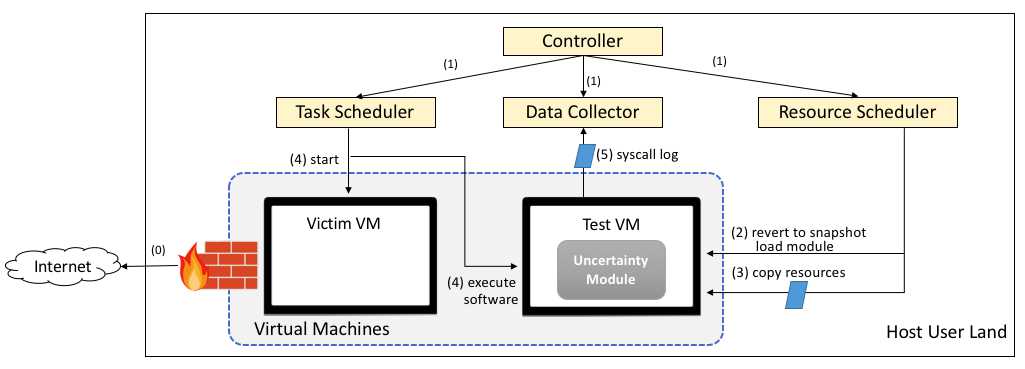}
    \vspace{-8pt}
	\caption{
		\footnotesize{The architecture of our evaluation testbed. The firewall restricts network from all VMs (Step 0). The Controller starts the 
		Resource Scheduler, the Task Scheduler and the Data Collector (Step 1). The Resource Scheduler reverts the Test VM to a clean 
		Snapshot and loads into it the uncertainty module (Step 2), copies the software resources (e.g. files and parameters needed) 
		(Step 3). The Task Scheduler starts the Dionaea service in the Victim VM and executes software in the Test VM (Step 4). The 
		Data Collector reads system call traces and execution results to aid crash analysis (Step 5).}}
	\label{fig:datagen}
\end{figure*}

We implemented a testbed to perform our experiments. The testbed is deployed in a host desktop running Ubuntu 14.04 with 16GB RAM, 160GB Hard Disk, x86\_64 architecture, and 8 processors. In each test, the testbed runs two VMs: (i) Test VM, running \chame and testing malware and benign software, and (ii) Victim VM, running with resources that the malware in the Test VM may want to attack. This subsection details the architecture of the testbed (Figure \ref{fig:datagen}) and the process of automating scalable experiments. 
It has two virtual machines and four other components running as user land processes: (i) a central \textit{Controller}, (ii) a \textit{Resource Scheduler}, (iii) a \textit{Task Scheduler} and (iv) a \textit{Data Collector}. 

%Our testbed is automated with four components working on the host server and four pools locating in the VMs (Figure~\ref{fig:datagen}). 
Before starting the experiments, we set up the firewall to block all the possible network connections between the host and the VMs for security purposes. The VMs can communicate with each other, and can reach outside network through port 53 (DNS) and port 80 (HTTP) needed for malware downloading payloads from the Internet (Step 0 in Figure \ref{fig:datagen}).

The \textit{Controller} is responsible for managing the other components. It first starts the \textit{Resource Scheduler} to prepare the files and parameters for all the experiments, then launches the \textit{Task Scheduler} to run malware and benign software in the Test VM, %As some malware need to attack another IP, we offer the victim slave pools for candidacy. 
and finally starts the \textit{Data Collector} to record system call traces and execution results of each experiment (Step 1).

The \textit{Resource Scheduler} is responsible for preparing the resources needed for each experiment in the Test VM. First, it reverts the Test VM into a clean snapshot---a preserved state of the VM that the user can return to repeatedly. 
% Our clean snapshot stores the state of a freshly booted system with no malware running on it. 
Then, it loads the uncertainty module to the Test VM (Step 2). Finally, it copies the software and the corresponding files and parameters needed during the execution from the host to the Test VM (Step 3). 

The \textit{Task Scheduler} is responsible for executing tasks in the VMs. This involves starting the Dionaea \cite{dionaea} service in the Victim VM, and executing malware or benign software in the Test VM (Step 4). The Dionaea \cite{dionaea} service provides the required resources, such as network services, for malware running in the Test VM. 

The \textit{Data Collector} is responsible for collecting system call traces logged in {\tt dmesg} and software execution results, including returned error and segmentation fault (Step 5).

\subsection{Security}
The goal of the security evaluation is to analyze the effects of \chame's uncertain environment in malware and benign software execution. 
% with a static perturbation threshold protected the system against several types of malware.  
We considered that a piece of malware was \textit{adversely affected} by the uncertain environment if it crashed or experienced issues in its executed. An execution is considered \textit{Crashed} if malware terminates before performing its malicious actions. An execution is considered \textit{Succeeded} if malware accomplished its intended tasks, such as injecting malicious payload into an executable. The following outcomes are non-exhaustive examples of hampered malware execution in the uncertain environment: (1) a virus that injects only part of the malicious code to an executable or source code file; (2) a botnet that loses commands sent to the bot herder; (3) a cracker that retrieves wrong or partial user credentials; (4) a spyware that redirects incomplete stdin, stdout or stderr of the victim; (5) a flooder that sends only a percentage of the total number of packets it attempted.

\begin{figure*}[!htb]
	\centering
	\includegraphics[width=0.7\linewidth]{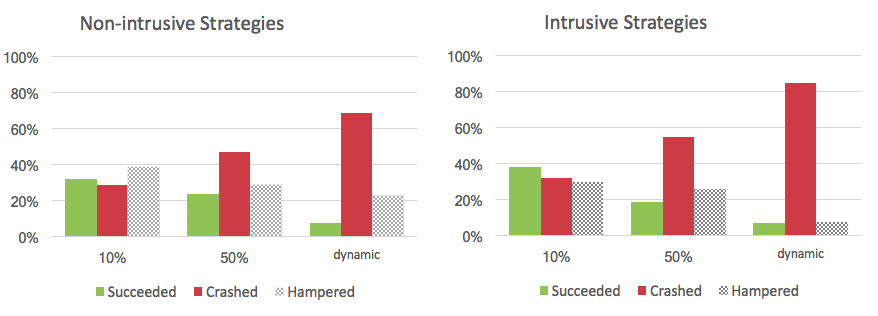}
    \vspace{-8pt}
	\caption{\footnotesize{Execution results for \textit{malware} running in the uncertain environment using intrusive and non-intrusive strategies 
	with static (10\% and 50\%) and per-system call thresholds.} }
	\label{fig:malware_res}
\end{figure*}

We evaluated the effects of the uncertain environment with 100 Linux malware samples (see Table 1  in the Appendix \cite{appendix} for the list) using intrusive and non-intrusive perturbation strategies and static and dynamic thresholds. As Figure \ref{fig:malware_res} shows, on average, intrusive strategies produced approximately 10\% more Crashed and 8\% fewer Hampered execution results than non-intrusive strategies. For both intrusive and non-intrusive strategies, the ratios of Succeeded malware execution (infection) were almost the same. When intrusive strategies were applied, 81\% of the malware samples failed to accomplish their tasks at threshold 50\%, 62\% failed at threshold 10\%, and 92\% failed with a dynamic per-system call threshold. Non-intrusive strategies yielded similar results for threshold 50\%, 10\% and per-system call threshold, with 76\%, 68\%, 93\% of malware adversely affected, respectively. In general, threshold 50\% caused more Crashed and fewer Succeeded malware execution results than threshold 10\%. The dynamic threshold caused 25\% fewer malware to succeed, and caused 30\% more malware to crash during execution than static threshold. This corroborates our assumption that a dynamic threshold focusing on likely malicious behavior will be effective at targeting malware.

We also ran our samples of general software in the uncertain environment and observed their execution outcome. Non-exhaustive examples of \textit{Hampered} executions are: (1) a text editor temporarily losing some functionality; (2) a scientific tool producing partial results; or (3) a network tool missing packets. The execution outcome was considered \textit{Crashed} if the software hanged longer than twice its standard runtime and needed to be manually killed. A \textit{Succeeded} execution generated outputs that matched those produced with the same test case in the standard environment and with a runtime that did not exceed twice than that in the standard runtime.

\begin{figure*}[!htb]
	\centering
	\includegraphics[width=0.7\linewidth]{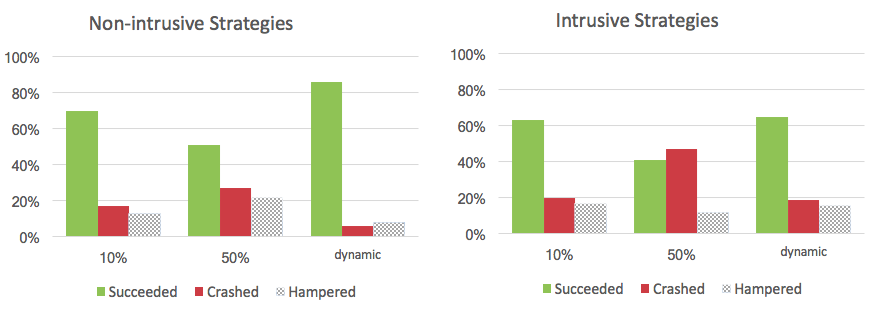}
% 	\vspace*{-\baselineskip}
	\caption{\footnotesize{Execution results for \textit{benign software} running in the uncertain environment using intrusive and non-intrusive 
	strategies with static (10\% and 50\%) and dynamic thresholds. }}
	\label{fig:benign_res}
	\vspace*{\baselineskip}
\end{figure*}

As expected (Figure \ref{fig:benign_res}), compared to non-intrusive strategies, intrusive strategies caused more adverse effects to benign software with approximately 10\% more Crashed, 7\% more Hampered, and 15\% fewer Succeed execution. At static threshold 10\% with intrusive strategies, on average, 37\% of the tasks experienced some form of Crashed or Hampered execution. With non-intrusive strategies, this percentage was 30\%. For a 50\% static threshold and intrusive strategies, 59\% of the software was adversely affected. With non-intrusive strategies, this number was 10\% smaller. A dynamic threshold with non-intrusive strategies on benign software led to 25\% more Succeeded, 15\% fewer Hampered, and 20\% fewer Crashed executions for benign software than a static threshold for the same configurations. With intrusive strategies, the effects of the dynamic threshold were similar to using a static threshold of 10\%. 

These results corroborate our hypothesis that the uncertain environment with a dynamic threshold can better isolate benign software from perturbations, and disproportionately affect malware compared to using static thresholds.

\subsection{Software Behavior and Performance}
We compared the execution of malware and benign software at the system call level in the uncertain environment. We explored the effects of different software types, software workloads, and perturbation strategies.

In our experiments, benign software invoked more than twice the number of system calls invoked by malware (even flooders, which usually invoke a large number of system calls). For benign software the number of system calls perturbed or silenced was only half of those of malware, mainly because of the effectiveness of the corruption protection mechanism (see Section \ref{sec:corrupt}). Benign software performed a larger number of connection attempts and read/write operations than malware. 

Table \ref{tab:mal-res-nonintr} and Table \ref{tab:mal-res-intr} show the results of system calls perturbed for different types of malware using non-intrusive and intrusive strategies. The impact of intrusive and non-intrusive strategies is similar.
Generally, threshold 50\% caused higher percentage of perturbations than a static threshold of 10\% and dynamic threshold, especially with connection-related system calls (50\% increase in the percentage of perturbations). The dynamic threshold caused higher percentage of perturbations than static threshold $t_d=10\%$, because a dynamic threshold could vary from $t_d=10\%$ to $t_{max}=95\%$. 
For all types of system calls invoked, flooders had the highest percentage of perturbations and worms had the lowest percentage of perturbations, with both static and dynamic threshold. Based on previous work \cite{sun2017dose}, this can be explained by flooders invoking a large number of system calls in the perturbation set, while worms invoking a smaller number. For network-related system calls, spyware had the lowest percentage of perturbation with both static and dynamic perturbation thresholds. With a dynamic threshold, spyware had 0\% of its network-related system calls perturbed. This can be explained that network-related system calls are usually invoked after {\tt sys\_dup(0)} system call, which was perturbed with threshold $t_{max}=95\%$. In other words, network-related system calls can hardly be invoked before spyware crashes. Similarly, for buffer-related system calls, spyware also experienced a very small number of perturbations, indicating a crash after {\tt sys\_dup(0)} system call.

\begin{table*}[!ht]
\centering
\scriptsize
\caption{\footnotesize{Percentage of system calls perturbed running malware with different thresholds in the uncertain environment using 
\textit{non-intrusive} strategies.}}
\label{tab:mal-res-nonintr}
\vspace*{-\baselineskip}
\begin{tabular}{|c|r|r|r|r|r|r|r|r|r|}
\hline
\multicolumn{1}{|l|}{} & \multicolumn{3}{c|}{\begin{tabular}[c]{@{}c@{}}Percentage of all syscalls \\ perturbed\end{tabular}} & \multicolumn{3}{c|}{\begin{tabular}[c]{@{}c@{}}Percentage of connection-related\\  calls perturbed\end{tabular}} & \multicolumn{3}{c|}{\begin{tabular}[c]{@{}c@{}}Percentage of buffer-related\\  calls perturbed\end{tabular}} \\ \cline{2-10} 
\multirow{-2}{*}{} & 10\% & \cellcolor[HTML]{C0C0C0}50\% & \cellcolor[HTML]{EFEFEF}Dynamic & 10\% & \cellcolor[HTML]{C0C0C0}50\% & \cellcolor[HTML]{EFEFEF}Dynamic & 10\% & \cellcolor[HTML]{C0C0C0}50\% & \cellcolor[HTML]{EFEFEF}Dynamic \\ \hline
Flooders & 9.74\% & \cellcolor[HTML]{C0C0C0}39.31\% & \cellcolor[HTML]{EFEFEF}37.91\% & 10.13\% & \cellcolor[HTML]{C0C0C0}59.29\% & \cellcolor[HTML]{EFEFEF}36.68\% & 6.58\% & \cellcolor[HTML]{C0C0C0}23.35\% & \cellcolor[HTML]{EFEFEF}24.92\% \\ \hline
Spyware & 2.89\% & \cellcolor[HTML]{C0C0C0}25.79\% & \cellcolor[HTML]{EFEFEF}14.33\% & 7.14\% & \cellcolor[HTML]{C0C0C0}48.15\% & \cellcolor[HTML]{EFEFEF}0.00\% & 3.06\% & \cellcolor[HTML]{C0C0C0}31.06\% & \cellcolor[HTML]{EFEFEF}0.41\% \\ \hline
Trojan & 8.09\% & \cellcolor[HTML]{C0C0C0}27.07\% & \cellcolor[HTML]{EFEFEF}21.17\% & 9.52\% & \cellcolor[HTML]{C0C0C0}62.14\% & \cellcolor[HTML]{EFEFEF}17.00\% & 7.14\% & \cellcolor[HTML]{C0C0C0}15.22\% & \cellcolor[HTML]{EFEFEF}1.81\% \\ \hline
Viruses & 5.02\% & \cellcolor[HTML]{C0C0C0}28.62\% & \cellcolor[HTML]{EFEFEF}23.47\% & 9.56\% & \cellcolor[HTML]{C0C0C0}47.78\% & \cellcolor[HTML]{EFEFEF}12.69\% & 4.96\% & \cellcolor[HTML]{C0C0C0}21.87\% & \cellcolor[HTML]{EFEFEF}17.27\% \\ \hline
Worms & 0.05\% & \cellcolor[HTML]{C0C0C0}15.67\% & \cellcolor[HTML]{EFEFEF}11.04\% & 9.86\% & \cellcolor[HTML]{C0C0C0}60.97\% & \cellcolor[HTML]{EFEFEF}8.11\% & 8.97\% & \cellcolor[HTML]{C0C0C0}14.37\% & \cellcolor[HTML]{EFEFEF}16.27\% \\ \hline
All & 0.41\% & \cellcolor[HTML]{C0C0C0}28.39\% & \cellcolor[HTML]{EFEFEF}19.80\% & 9.87\% & \cellcolor[HTML]{C0C0C0}60.97\% & \cellcolor[HTML]{EFEFEF}15.69\% & 6.83\% & \cellcolor[HTML]{C0C0C0}21.10\% & \cellcolor[HTML]{EFEFEF}13.81\% \\ \hline
\end{tabular}
\vspace*{\baselineskip}
\end{table*}

\begin{table*}[!ht]
\centering
\scriptsize
\caption{\footnotesize{Percentage of system calls perturbed running malware with different thresholds in the uncertain environment using 
\textit{intrusive} strategies.}}
\label{tab:mal-res-intr}
\vspace*{-\baselineskip}
\begin{tabular}{|c|r|r|r|r|r|r|r|r|r|}
\hline
\multicolumn{1}{|l|}{} & \multicolumn{3}{c|}{\begin{tabular}[c]{@{}c@{}}Percentage of all syscalls \\ perturbed\end{tabular}} & \multicolumn{3}{c|}{\begin{tabular}[c]{@{}c@{}}Percentage of connection-related\\  calls perturbed\end{tabular}} & \multicolumn{3}{c|}{\begin{tabular}[c]{@{}c@{}}Percentage of buffer-related\\  calls perturbed\end{tabular}} \\ \cline{2-10} 
\multicolumn{1}{|l|}{\multirow{-2}{*}{}} & 10\% & \cellcolor[HTML]{C0C0C0}50\% & \cellcolor[HTML]{EFEFEF}Dynamic & 10\% & \cellcolor[HTML]{C0C0C0}{\color[HTML]{333333} 50\%} & \cellcolor[HTML]{EFEFEF}Dynamic & 10\% & \cellcolor[HTML]{C0C0C0}50\% & \cellcolor[HTML]{EFEFEF}Dynamic \\ \hline
Flooders & 8.22\% & \cellcolor[HTML]{C0C0C0}39.28\% & \cellcolor[HTML]{EFEFEF}39.11\% & 9.56\% & \cellcolor[HTML]{C0C0C0}{\color[HTML]{333333} 49.57\%} & \cellcolor[HTML]{EFEFEF}38.37\% & 3.50\% & \cellcolor[HTML]{C0C0C0}23.11\% & \cellcolor[HTML]{EFEFEF}30.69\% \\ \hline
Spyware & 4.38\% & \cellcolor[HTML]{C0C0C0}26.39\% & \cellcolor[HTML]{EFEFEF}15.02\% & 16.62\% & \cellcolor[HTML]{C0C0C0}{\color[HTML]{333333} 51.25\%} & \cellcolor[HTML]{EFEFEF}37.96\% & 0.64\% & \cellcolor[HTML]{C0C0C0}27.14\% & \cellcolor[HTML]{EFEFEF}0.07\% \\ \hline
Trojan & 6.90\% & \cellcolor[HTML]{C0C0C0}35.16\% & \cellcolor[HTML]{EFEFEF}25.08\% & 12.49\% & \cellcolor[HTML]{C0C0C0}{\color[HTML]{333333} 56.62\%} & \cellcolor[HTML]{EFEFEF}19.45\% & 3.58\% & \cellcolor[HTML]{C0C0C0}27.47\% & \cellcolor[HTML]{EFEFEF}6.08\% \\ \hline
Viruses & 6.49\% & \cellcolor[HTML]{C0C0C0}23.03\% & \cellcolor[HTML]{EFEFEF}28.14\% & 12.96\% & \cellcolor[HTML]{C0C0C0}{\color[HTML]{333333} 52.34\%} & \cellcolor[HTML]{EFEFEF}14.30\% & 8.94\% & \cellcolor[HTML]{C0C0C0}22.32\% & \cellcolor[HTML]{EFEFEF}24.19\% \\ \hline
Worms & 3.92\% & \cellcolor[HTML]{C0C0C0}22.93\% & \cellcolor[HTML]{EFEFEF}13.26\% & 6.53\% & \cellcolor[HTML]{C0C0C0}{\color[HTML]{333333} 60.59\%} & \cellcolor[HTML]{EFEFEF}8.15\% & 3.52\% & \cellcolor[HTML]{C0C0C0}19.86\% & \cellcolor[HTML]{EFEFEF}33.24\% \\ \hline
All & 6.26\% & \cellcolor[HTML]{C0C0C0}29.90\% & \cellcolor[HTML]{EFEFEF}26.27\% & 11.26\% & \cellcolor[HTML]{C0C0C0}{\color[HTML]{333333} 53.55\%} & \cellcolor[HTML]{EFEFEF}26.79\% & 4.47\% & \cellcolor[HTML]{C0C0C0}24.04\% & \cellcolor[HTML]{EFEFEF}19.36\% \\ \hline
\end{tabular}
\vspace*{\baselineskip}
\end{table*}

Tables \ref{tab:ben-res-nonintr} and \ref{tab:ben-res-intr} show the results of system call perturbation for benign I/O-bound and CPU-bound software using non-intrusive and intrusive strategies. In general, a threshold of 50\% caused higher percentages of system calls perturbed than a threshold of 10\%. With static thresholds of 10\% and 50\%, compared with IO-bound software, CPU-bound software experienced a higher percentage of system calls perturbed, and lower percentages of connection-related system calls and buffer-related system calls perturbed. 
Dynamic threshold applied to I/O-bound software caused a higher percentage of system calls perturbed compared to CPU-bound software, mainly because the dynamic threshold is likely to affect IO-related system calls more (see Behaviors in Section \ref{sec:threshold}).

\begin{table*}[!ht]
\centering
\scriptsize
\caption{\footnotesize{Percentage of system calls perturbed running benign software with different thresholds in the uncertain environment using 
\textit{non-intrusive} strategies.}}
\label{tab:ben-res-nonintr}
\vspace*{-\baselineskip}
\begin{tabular}{|c|r|r|r|r|r|r|r|r|r|}
\hline
\multicolumn{1}{|l|}{} & \multicolumn{3}{c|}{\begin{tabular}[c]{@{}c@{}}Percentage of all syscalls \\ perturbed\end{tabular}} & \multicolumn{3}{c|}{\begin{tabular}[c]{@{}c@{}}Percentage of connection-related\\  calls perturbed\end{tabular}} & \multicolumn{3}{c|}{\begin{tabular}[c]{@{}c@{}}Percentage of buffer-related\\  calls perturbed\end{tabular}} \\ \cline{2-10} 
\multicolumn{1}{|l|}{\multirow{-2}{*}{}} & 10\% & \cellcolor[HTML]{C0C0C0}50\% & \cellcolor[HTML]{EFEFEF}dynamic & 10\% & \cellcolor[HTML]{C0C0C0}50\% & \cellcolor[HTML]{EFEFEF}dynamic & 10\% & \cellcolor[HTML]{C0C0C0}50\% & \cellcolor[HTML]{EFEFEF}dynamic \\ \hline
IO & 1.24\% & \cellcolor[HTML]{C0C0C0}7.65\% & \cellcolor[HTML]{EFEFEF}2.60\% & 5.94\% & \cellcolor[HTML]{C0C0C0}23.40\% & \cellcolor[HTML]{EFEFEF}2.50\% & 0.94\% & \cellcolor[HTML]{C0C0C0}6.34\% & \cellcolor[HTML]{EFEFEF}34.16\% \\ \hline
CPU & 3.42\% & \cellcolor[HTML]{C0C0C0}10.72\% & \cellcolor[HTML]{EFEFEF}0.10\% & 0.00\% & \cellcolor[HTML]{C0C0C0}1.97\% & \cellcolor[HTML]{EFEFEF}0.00 & 3.39\% & \cellcolor[HTML]{C0C0C0}10.16\% & \cellcolor[HTML]{EFEFEF}0.02\% \\ \hline
All & 2.40\% & \cellcolor[HTML]{C0C0C0}9.28\% & \cellcolor[HTML]{EFEFEF}1.28\% & 2.79\% & \cellcolor[HTML]{C0C0C0}12.04\% & \cellcolor[HTML]{EFEFEF}1.18\% & 2.24\% & \cellcolor[HTML]{C0C0C0}8.36\% & \cellcolor[HTML]{EFEFEF}15.99\% \\ \hline
\end{tabular}
\vspace*{\baselineskip}
\end{table*}

\begin{table*}[!ht]
\centering
\scriptsize
\caption{\footnotesize{Percentage of system calls perturbed running benign software with different thresholds in the uncertain environment using 
\textit{intrusive} strategies.}}
\label{tab:ben-res-intr}
\vspace*{-\baselineskip}
\begin{tabular}{|c|r|r|r|r|r|r|r|r|r|}
\hline
\multicolumn{1}{|l|}{} & \multicolumn{3}{c|}{\begin{tabular}[c]{@{}c@{}}Percentage of all syscalls \\ perturbed\end{tabular}} & \multicolumn{3}{c|}{\begin{tabular}[c]{@{}c@{}}Percentage of connection-related\\  calls perturbed\end{tabular}} & \multicolumn{3}{c|}{\begin{tabular}[c]{@{}c@{}}Percentage of buffer-related\\  calls perturbed\end{tabular}} \\ \cline{2-10} 
\multirow{-2}{*}{} & 10\% & \cellcolor[HTML]{C0C0C0}50\% & \cellcolor[HTML]{EFEFEF}dynamic & 10\% & \cellcolor[HTML]{C0C0C0}50\% & \cellcolor[HTML]{EFEFEF}dynamic & 10\% & \cellcolor[HTML]{C0C0C0}50\% & \cellcolor[HTML]{EFEFEF}dynamic \\ \hline
IO & 1.21\% & \cellcolor[HTML]{C0C0C0}5.62\% & \cellcolor[HTML]{EFEFEF}8.20\% & 3.57\% & \cellcolor[HTML]{C0C0C0}20.14\% & \cellcolor[HTML]{EFEFEF}2.50\% & 0.95\% & \cellcolor[HTML]{C0C0C0}3.88\% & \cellcolor[HTML]{EFEFEF}50.93\% \\ \hline
CPU & 3.34\% & \cellcolor[HTML]{C0C0C0}11.82\% & \cellcolor[HTML]{EFEFEF}2.40\% & 0.03\% & \cellcolor[HTML]{C0C0C0}2.01\% & \cellcolor[HTML]{EFEFEF}1.00\% & 3.49\% & \cellcolor[HTML]{C0C0C0}12.27\% & \cellcolor[HTML]{EFEFEF}8.10\% \\ \hline
All & 2.34\% & \cellcolor[HTML]{C0C0C0}8.91\% & \cellcolor[HTML]{EFEFEF}5.13\% & 1.69\% & \cellcolor[HTML]{C0C0C0}10.53\% & \cellcolor[HTML]{EFEFEF}1.71\% & 2.30\% & \cellcolor[HTML]{C0C0C0}8.33\% & \cellcolor[HTML]{EFEFEF}28.21\% \\ \hline
\end{tabular}
\vspace*{\baselineskip}
\end{table*}

One of the greatest differences between the malware and benign software analyzed in this study is the diversity of the latter. To ensure a fair analysis of benign software, we measured the test coverage (percentage of software instructions executed) by compiling benign software source code with gcov \cite{gcov}, EMMA \cite{emma}, and Coverage.py \cite{coveragepy} based on the software's programming language. The average coverage for benign software in our analysis was 69.49\%. 

\begin{figure}[!ht]
\centering
\includegraphics[width=0.35\textwidth,trim={8 8 8 8},clip]{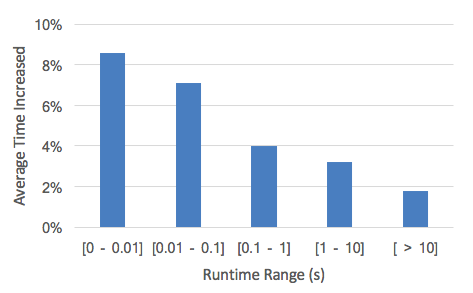}%[width=0.45\textwidth]
\vspace{-8pt}
\caption{\footnotesize{Performance penalty for 23 benchmark software whose execution time could be scripted. We categorized the software according to 
their average runtime.}}
\label{fig:runtime} %label needs to go after caption
\end{figure}

We also analyzed the performance penalty caused by the perturbation strategies, such as process delay and process priority decrease on all 23 benchmark software whose execution could be scripted (see Table 2 in the Appendix \cite{appendix} for a list of these benchmark software). Highly interactive software was tested manually and showed negligible performance overhead. Figure \ref{fig:runtime} shows the average runtime overhead for software whose execution could be scripted running in the uncertain environment. For runtimes ranging from 0 to 0.01 seconds, the average penalty was 8\%; for runtimes ranging from 0.1 to 1 seconds, the average penalty was 4\%; for runtimes longer than 10 seconds, the average penalty was 1.8\%. This shows that the longer the runtime, the smaller the performance overhead. One hypothesis is that software with longer execution time is usually CPU-bound. Because most of the system calls in the perturbation set are I/O related, CPU-bound programs end up being perturbed less intensively.

We also tested 26 benign applications with different workloads running in the standard and uncertain environment (see Table 2 in the Appendix \cite{appendix} for the list of these applications). The workloads were characterized under three levels: \textit{light}, \textit{medium} and \textit{heavy}, which corresponded to \textit{test}, \textit{train}, and \textit{ref} level for SPEC CPU2006, and first, middle-most, and last-level in the Phoronix Test Suite. %In the uncertain environment, we tested all three workloads for each of the 26 benign software with the intrusive and non-intrusive strategies.
On all three different workloads, our results showed that two benign software were adversely affected by non-intrusive strategies and nine software were affected by intrusive strategies (see Table \ref{tab:workload}). Further, there were no significant changes on the percentages of total system calls perturbed, connection-related system calls perturbed and buffer-related system calls perturbed with the change of workloads for both types of perturbation strategies. The results indicate that the workload type of the tested software does not impact the program outcome in the uncertain environment for the two sets of perturbation strategies we used.

\begin{table*}[!ht]
\centering
\scriptsize
\caption{\footnotesize{Impact of non-intrusive and intrusive strategies on 26 benign software from Phoronix Test Suite and SPEC CPU for different workloads in the uncertain environment (static threshold 10\%). %We computed the byte loss as the number of bytes received as parameter in file-system related system calls.
}}
\label{tab:workload}
\vspace*{-\baselineskip}
\begin{tabular}{|c|c|c|c|c|c|c|c|c|}
\hline 
 & \multicolumn{2}{c|}{\begin{tabular}[c]{@{}c@{}}Percentage of \\ syscalls perturbed\end{tabular}} & \multicolumn{2}{c|}{\begin{tabular}[c]{@{}c@{}}Percentage of \\ connection-related \\ syscalls perturbed\end{tabular}} & \multicolumn{2}{c|}{\begin{tabular}[c]{@{}c@{}}Percentage of \\ buffer-related \\ syscalls perturbed\end{tabular}} & \multicolumn{2}{c|}{\begin{tabular}[c]{@{}c@{}}Number of \\ Crashes\end{tabular}} \\ \cline{2-9} 
\multirow{-2}{*}{Workload} & Non-intrusive & \cellcolor[HTML]{EFEFEF}Intrusive & Non-intrusive & \cellcolor[HTML]{EFEFEF}Intrusive & Non-intrusive & \cellcolor[HTML]{EFEFEF}Intrusive & Non-intrusive & \cellcolor[HTML]{EFEFEF}Intrusive \\ \hline 
Light & 4.3\% & \cellcolor[HTML]{EFEFEF}5.1\% & 0.0\% & \cellcolor[HTML]{EFEFEF}0.1\% & 2.9\% & \cellcolor[HTML]{EFEFEF}4.2\% & 2 & \cellcolor[HTML]{EFEFEF}9 \\ \hline
Medium & 5.8\% & \cellcolor[HTML]{EFEFEF}6.3\% & 0.2\% & \cellcolor[HTML]{EFEFEF}0.3\% & 3.1\% & \cellcolor[HTML]{EFEFEF}3.7\% & 2 & \cellcolor[HTML]{EFEFEF}9 \\ \hline
Heavy & 5.2\% & \cellcolor[HTML]{EFEFEF}5.9\% & 0.2\% & \cellcolor[HTML]{EFEFEF}0.2\% & 3.5\% & \cellcolor[HTML]{EFEFEF}3.0\% & 2 & \cellcolor[HTML]{EFEFEF}9 \\ \hline
\end{tabular}

\end{table*}

\subsection{Effects of Uncertainty on Application Execution}
In this section, we describe how 
\chame can leverage random perturbations to thwart malware 
samples, from poorly-programmed to evasive ones, and also to
aid the discovery of bugs in benign software.

\subsubsection{Effects of Uncertainty on Malware}

\noindent \textbf{Black Vine} 
We simulated a watering hole attack similar to the \textit{Black Vine} malware from Symantec \cite{blackvine}. This attack has three main components: a Trojan, a backdoor and a keylogger. 
%Figure \ref{fig:apt} shows the workflow of this simulated APT. 
First, the attacker sends a phishing e-mail to a user with a link for downloading the Trojan encryption tool. If the user clicks on the link and later uses the Trojan to encrypt a file, the tool downloads and executes a backdoor from a C\&C server while encrypting the requested file. Then, the backdoor copies the directory structure and the ssh host key from the user's machine into a file and sends it to the C\&C server. 
After the backdoor executes, the attacker deletes any traces of the infection without affecting the Trojan's encryption/decryption functionality. The attacker will also install a keylogger to obtain root privileges. Next, the backdoor runs a script that uploads sensitive data to the C\&C server.

The Trojan is written in C using \textit{libgcrypt} for encryption and decryption. It uses the curl library for downloading the backdoor from the Internet. In our simulation we used the \textit{logkeys} keylogger \cite{logkeys}. The backdoor script uses scp for sending the data to the C\&C server. 

From the system call traces we collected, the first malicious behavior occurred when the backdoor was being configured, with a {\tt sys\_write()} invoked with a buffer parameter starting with $\backslash$177ELF. %, and it has a probability $t_{1}$ being failed. 
This behavior caused the threshold to increase to $t_\text{A1}$ on the {\tt sys\_write()} system call. Later, three pairs of {\tt sys\_dup2()} with file descriptors 0 and 1 are invoked afterwards to execute the backdoor. The threshold on the three {\tt sys\_dup2()} was increased to $t_\text{A2}$.
Then, when {\tt sys\_read()} on the ssh host key files was invoked, the threshold decreased to $t_\text{d}$. Finally, the keylogger started, {\tt sys\_write()} was invoked to write to a log file and {\tt sys\_connect()} and {\tt sys\_sendto()} were invoked for the backdoor to communicate with the C\&C server. The probability for the simulated malware to gain privilege and exfiltrate data is under $(1-t_\text{A1})\times (1-t_\text{A2})^{3}\times (1-t_\text{d}\%)$, which is 0.14\%. In our 15 experiments, nine crashed before setting up the backdoor, four crashed before starting the keylogger, and two crashed before communicating with the C\&C server.

\noindent \textbf{Poorly-written Malware}. 
We also evaluated a poorly-programmed malware sample~\cite{distributed}, 
% (a variation from the malware described by B. Prince at~\cite{distributed}), 
which splits its infection operation into two threads: the first (Code~\ref{lst:t1}) gets access permissions to a given directory whereas the second waits for access permissions to infect the files in the directory (Code~\ref{lst:t2}).

\begin{minipage}[t]{0.45\columnwidth}
\begin{lstlisting}[label=lst:t1,caption=Thread 1]
get_permissions();
finish();
\end{lstlisting}
\end{minipage}
\begin{minipage}[t]{0.45\columnwidth}
\begin{lstlisting}[label=lst:t2,caption=Thread 2.]
sleep(TIME);
infect();
\end{lstlisting}
\end{minipage}

The threads are synchronized using a sleep call instead
of a system lock, which can be considered a poor programming practice, because different systems might have different sleep times. 
While running in \chame's uncertain environment,
unexpected sleep responses may wake up \texttt{Thread 2}
before accesses permissions are granted by 
\texttt{Thread 1}, thus causing the malware to fail. In our 15 tests, 
the sample failed 12 times because of the short sleep time (\chame's perturbations caused a decrease in the sleep time). In the three times the sample got permissions, it was mitigated by the increased perturbation threshold due to the frequent invocation of a system call (the sample iterated the \texttt{/proc} folder 
to find process identifiers).

\noindent\textbf{Evasive Samples}
Advanced malware use a number of evasive techniques 
to avoid being detected while executing in sandboxes 
and/or antivirus (AV) emulators. A popular evasion 
method is execution stalling, i.e., malware postpones 
its malicious behavior until the sandbox times
out~\cite{evasivemalware}. Execution stalling
techniques were identified in up to 64\% of all 
Linux samples considered in a longitudinal
study~\cite{cozzi2018understanding}. Malware 
samples stall their executions by employing 
distinct techniques, such as:

\noindent 1) \textit{Stalling Strategy 1}: malware performs a 
configurable and random long sleep before and during 
exhibiting its malicious behavior aiming to trigger
sandbox execution timeouts.\\
2) \textit{Stalling Strategy 2}: malware performs a great number of memory allocations (e.g. {\tt malloc}) to slow down its operation while running under an emulator that traps memory accesses.\\
3) \textit{Stalling Strategy 3}: malware delays its execution by performing long and costly computations, such as float point-based operations---as emulating float-based operations is hard and often results in sandboxes timeouts.

\begin{table}[!ht]
\centering
\scriptsize
\caption{\footnotesize{Malware with evasive techniques running in the standard OS and in \chame.}}
\label{tab:evasive}
\vspace*{-\baselineskip}
\begin{tabular}{|c|c|c|c|}
\hline
\begin{tabular}[c]{@{}c@{}}Evasive\\ Strategies\end{tabular} & \begin{tabular}[c]{@{}c@{}}Percentage\\of syscalls\\perturbed\end{tabular} & \begin{tabular}[c]{@{}c@{}}Percentage \\of buffer bytes\\perturbed\end{tabular} & \begin{tabular}[c]{@{}c@{}}Execution results \\in \chame \end{tabular}\\ \hline 
Stalling 1 & 3.07\% & 0.07\% & Operation not permitted\\ \hline
Stalling 2 & 2.31\% & 0.68\% & Segmentation Fault\\ \hline
Stalling 3 & 3.02\% & 0.46\% & Operation not permitted\\ \hline
\end{tabular}
\end{table}

We studied stalling tactics by armoring a malware sample with these three evasive techniques independently investigated how \chame 
thwarts their operations. The malware sample was chosen as a challenging case because it would not set up any backdoor, nor would it inject malicious payloads to other executables, and most of the files the sample accessed were under the corruption protection mechanism. The sample would: i) fingerprint the environment 
to get an infection identifier; ii) install a proxy to infect the 
browser; iii) infect the DNS subsystem modifying the default routes; 
and iv) finally remove all the logs generated. The analyzed sample 
would not set up any backdoor, nor would it inject malicious payloads to 
other executables.

We ran these three different scenarios using the three different stalling techniques in both the standard environment and under \chame. In the standard environment, the malware sample could effectively exhibit its malicious behavior under all three scenarios.   
In \chame, configured with a dynamic threshold and non-intrusive strategies, the sample failed to achieve its goal. We measured how each stalling strategy was perturbed and summarized the results in Table \ref{tab:evasive}.
\chame silenced approximately 3\% of system calls in the perturbation set, and perturbed less than 1\% of buffer bytes. The percentage is small because of the corruption protection mechanism, which prevented many perturbations. 
For all three evasive scenarios, \chame mitigated the malware through either crash or segmentation fault. With the \texttt{Stalling Strategy 1} and \texttt{Stalling Strategy 3}, the malware sample terminated earlier (compared to executions in the standard environment), due to the perturbation on the {\tt read} system call. In the user space, the malware sample was deceived with an error message ``Operation not permitted in accessing file". With \texttt{Stalling Strategy 2}, the malware sample terminated with a segmentation fault due to the perturbation on {\tt mmap} system call. The malware saw the memory allocation failure and terminated without executing malicious behavior, as the malware assumed it was running under an emulator. 
Notice that these samples successfully executed outside
of \chame environment, thus the imposed perturbations are
the only responsible for their unsuccessful execution. The
OS-related error messages are due to \chame's perturbations
messing with malware internal states, thus resulting in
invalid requests.

\subsection{\chame evolves over time}
Since \chame has demonstrated its potential in malware mitigation, we further discuss whether \chame can stand the test of time by extending the malware sample set and testing them on newer version systems. The extended set contains 5,679 Linux malware samples collected from Virustotal \cite{virustotal} and VirusShare \cite{virusshare}. In this set, 273 samples can successfully attack our new testing systems, running Linux Ubuntu 4.15.0-20-generic version (Ubuntu 18.04). 

We carried out the experiments with per-system call threshold and randomly pick non-intrusive and intrusive strategies. The new results showed that 89.38\% of the malware samples are mitigated, with 43.59\% of the malware hampered and 45.79\% crashed. Compared with prior experiments on Ubuntu 12.04 (six years before the new testing system), which mitigated around 90\% of the malware (Figure \ref{fig:malware_res}), the new results corroborated the old ones. It is also safe to say that \chame can evolve with malware over time.

Further, we measured the detection rates of existing Anti-virus software on this extended set through Virustotal. The top 10 highest detection rates (ranging from 65\% -85\%) with their company names are listed in Table \ref{tab:avs}.

\begin{table}[!ht]
\scriptsize
\caption{\footnotesize{\textbf{TOP-10 AV Detection.} The best
rate of 85\% is ssmaller than the achieved by \chame.}}
\label{tab:avs}
\vspace*{-\baselineskip}
\begin{tabular}{cc|cc}
\hline
\textbf{AV} & \textbf{Detection (\%)} & \textbf{AV}          & \textbf{Detection (\%)} \\ \hline
Ikarus      & 85.61\%                 & AVG                  & 70.58\%                 \\
Kaspersky   & 79.83\%                 & Symantec             & 69.80\%                 \\
ESET-NOD32  & 78.81\%                 & TrendMicro-HouseCall & 68.61\%                 \\
GData       & 73.16\%                 & Sophos               & 67.96\%                 \\
Avast       & 72.99\%                 & Qihoo-360            & 64.37\%                \\ \hline
\end{tabular}
\end{table}

\subsubsection{Effects of Uncertainty on Benign Software}
\begin{table}[!ht]
\centering
\scriptsize
\caption{\footnotesize{Software bugs found by \chame}}
\label{tab:bug}
\vspace*{-\baselineskip}
\begin{tabular}{|c|l|}
\hline
\textbf{Software} & \multicolumn{1}{c|}{\textbf{Bugs}} \\ \hline 
Vim & viminfo: Illegal starting char \cite{viminfobug} \\ \hline
\multirow{2}{*}{tar} & \begin{tabular}[c]{@{}l@{}}Fail using '-C' option extracting archive with empty \\ directories \cite{taremptybug}\end{tabular}\\  \cline{2-2}
 & ``Operation not permitted" when extracting \cite{taropbug}\\ \hline
\multirow{2}{*}{Thunderbird} & Unable to locate mail spool file \cite{thunderspool}\\ \cline{2-2} 
 & segmentation fault (core dumped) \cite{thunderseg}\\ \hline
\multirow{2}{*}{Firefox} & Bus error (core dumped) \cite{firefoxbus} \\ \cline{2-2} 
 & Fatal IO error (Operation not permitted) on X server \cite{firefoxIO}\\  \hline
\end{tabular}
\end{table}

\chame's perturbations also affect benign applications. Whereas 
we observed that common benign applications can handle 
the unexpected system call responses in most of all 
experiments, we identified some crashes during their execution.
Therefore, to better understand \chame's impact and to 
identify how benign software could improve to 
better adapt to the perturbations, we \textit{manually} inspected
all crashed execution traces.

We analyzed the execution logs, from the last system call 
executed (including its parameters) until the first system call
executed in a reversed order. We observed that usually the failure of one system call with a specific parameter would lead to application early termination. Therefore, locating the corresponding system call causing the crash and its corresponding parameters would reveal the reason for the crash, and could potentially help the process of finding bugs. 
\chame is capable of perturbing every system call with a probability (given by the perturbation threshold), and logging the execution details (perturbations, system calls invoked and their parameters) about the crash. 

During our analysis, we found that the crashes in Vim, tar, Mozilla Firefox and Thunderbird were in fact software bugs previously reported on Launchpad and Bugzilla \cite{launchpad,bugzilla}. Because each system call was perturbed with a probability, the perturbations causing the crash in different tests varied on the same software (we ran each software fifteen times and averaged the results). Therefore, different bugs could be found for one piece of software. Table \ref{tab:bug} lists the bugs in detail. Besides general bugs, e.g. Segmentation Fault, Fatal I/O error and Bus error, we found several bugs of particular interest. 

\textbf{.viminfo} \cite{viminfobug}: This bug causes Vim to fail to launch because of an erroneous .viminfo file. The .viminfo file is used to record information about the last edits from a user. If the user exits Vim and later re-starts it, the .viminfo file enables the user to continue where he left it off \cite{viminfodoc}. In our experiment, the .viminfo error was caused by silencing a {\tt sys\_write} on .viminfo file. In the reported bug, the error was caused by an operation using a special character not recognized by Vim before exiting. With \chame, we identified that the reason for Vim stopping launching is the failure of {\tt sys\_open} on .viminfo. This shows the lack of fail-safe defaults from Vim.

\textbf{tar -C empty directory} \cite{taremptybug}: This bug occurs when one extracts empty directories inside an archive using the `-C' option to change directories. The cause for the bug is tar using {\tt mkdir (file\_name, mode)} instead of {\tt mkdirat (chdir\_fd, file\_name, mode)} to extract a directory. With \chame, we identified the failure of creating a new file descriptor with {\tt sys\_openat} in our log file,  showing that tar currently do not handle failures on that particular invocation of the system call well.

\textbf{Thunderbird mail spool file} \cite{thunderspool}: The bug causes Thunderbird to hang when linking an existing email account.
Thunderbird uses the spool file to ``help" the user set up an email account with the assumption that the email providers set up SMTP, ports, and security configurations very well. Unfortunately, few of them are correctly configured \cite{thunderbirdspool}. From the log files of \chame, we identified that the failure in linking an account was caused by a failure in {\tt sys\_read} of spool file. 

Our results show that the crashes and adverse effects in the analyzed benign software were in many cases actually caused by bugs (previously reported) instead of the perturbations applied by \chame. It appears that the perturbations just accelerated the exposure of such bugs, thus, showing that \chame could be also potentially applied to test software reliability.

\section{Discussion}\label{sec:discuss}

This section will discuss about the insights and limitations based on \chame's findings. 

\subsection{Defensive Programming}
The adoption of defensive programming for malware writing increases
its development costs both in human and in time resources. The paradigm of 
defensive programming is based on the assumption that runtime errors in
software are going to arise for a variety of reasons (including a potentially malicious OS) and that software needs to be written 
so as to be resilient to such errors~\cite{swc-osg-workshop}. This 
requires software developers (or malware writers for that matter) to 
include assertions in programs at runtime, and to write tests suites that anticipate different error scenarios and discover unknown behaviors.

Upon encountering invalid inputs, it is advantageous for malware to fail 
early and quickly to prevent leaving fingerprint of 
their actions in the system. Therefore, it is plausible to hypothesize that, under these
conditions, malware writers would need to write more reliable code so as to successfully operate under runtime uncertain conditions. 

Currently, a great number of malware source code found in the wild exhibit poor programming practices, such as lack of error checking
routines. Even more noticeable, many malware implementations 
are flawed, as seen on numerous ransomware decryption keys
retrieved via exploitation of flawed implemented 
routines~\cite{ransom,ransom2,ransom3,ransom4,ransom5}.

We acknowledge that the adoption of a paradigm such as \chame could further incentivize
malware writers into adopting defensive programming and,
consequently, writing ``resilient'' malware. However, this would come
under higher costs for malware writers.

Defensive programming might seem difficult to grasp in the beginning, like any new concept in software development. In fact, this concept has been well established and many programming languages, such as D, have already supported pre- and post-conditions as fundamental parts of their syntax \cite{depth_defensive_prog}. With the increasing need of defensive programming, programmers and testers can work on developing automated test suites to anticipate different error scenarios and discover unknown behaviors. In the era of cloud computing combined with the Internet of Things (IoT) and Artificial Intelligence (AI), more and more small services will interact with one another dynamically, making defensive programming essential to ensure composed system reliability. We hypothesize that defensive programming will become built-in design characteristics of software development.

\subsection{Trade-off: Performance vs. Mitigation Effect}
\textbf{Strategy}: There are trade-offs in selecting a perturbation strategy. Non-intrusive perturbation strategies that delay system calls, decrease the priority, or silence the system calls with error returns, aim to slow down program execution, potentially buying time for a deep learning model to operate. End users might experience a system slowdown in exchange for more security. Intrusive perturbation strategies, which are more aggressive, are designed for organizations with higher security expectations.
% and using defensive programming as a software development requirement. 
% For example, NASA requires applications to declare objects at the smallest possible level of scope, check the return value of non-void functions, and use static and dynamic assertions as sanity checks \cite{nasa_defensive_prog}.

% However, our approach is not suitable for organizations that do not control software running in their perimeter. 

\textit{Process Delay} strategy is different from suspending software execution. A suspended execution stops suspicious software from running and would not generate data for a potential deep learning analysis. \textit{Process Delay}, actually, slows down software execution, potentially buying time for deep analysis and allowing for a more accurate classification of software which received borderline confidence levels in classifications by a fast conventional machine learning detector. Moreover, suspension of execution can be detected by malware just by checking wall clock time. 
% this was written by Matt Bishop

\textbf{Threshold}: We acknowledge that the degree of uncertainty is not a one-size-fits-all solution. Based on the needs of the organization and its applications, \chame's perturbation threshold can be adjusted by a system administrator according to the organization's expectations and requirements. Based on our manual analysis, an acceptable static threshold is no more than 50\% and a dynamic threshold is no more than 95\%. 
The perturbation threshold can also be automatically adjusted following security policies. For example, \chame can raise the threshold for a given application if it passes a round of deep learning classification. 

Moreover, \chame can leverage an attribution-based scheme for initial threshold assignment. If the software has no attribution of origin, no supply chain, or its origin is not trusted, \chame can set a higher default threshold for the application. Conversely, if the application is trusted, \chame can set a lower default threshold for it. We leave such approaches for future work.

\textbf{Comparison with existing AV}: 
As Table \ref{tab:avs} shows, the best performed AV software produces a smaller detection rate than \chame. In addition, \chame complements existing AV software, such as FireEye \cite{fireeye}, whose goals are either to identify the signature or monitor the runtime behavior through isolating the software for a while until a decision can be made. However, some software may not exhibit its malicious behavior at the very beginning, or confuse the AV software by demonstrating benign behaviors for most of the time. 

\chame enables lifelong execution for software whose behaviors are hard to define and which cause borderline classification decisions. During the software execution, suspicious software will be placed in the uncertain environment when some borderline malicious behavior is detected, and be transferred between the standard environment and the uncertain environment multiple times.

\textbf{Limitations}: \chame is limited in mitigating well-written malware. Highly fault-tolerant malware will have a chance to succeed \chame. Further, if an administrator decides to whitelist common benign software from \chame, in-memory-only attacks \cite{faros} that inject themselves in the address space of a benign software will also succeed.

\subsection{Linux vs. Windows}
\textbf{Linux Malware}: 
Linux promotes open source code and streamlines
prototyping. This makes \chame to be deployed in Linux because of its resources in kernel programming. Linux also makes it easier for manual analysis at the beginning to understand the malware behavior
However, finding Linux-based malware
samples is challenging. First, the
availability of Linux malware samples is 
more restricted in comparison to other
platforms, such as Windows. For instance, 
the Cozzi et al's 
work~\cite{cozzi2018understanding},
the largest and most current Linux 
malware study, leveraged 10,000 malware 
samples whereas Windows studies may 
encompass million 
samples~\cite{hardy2016dl4md}. Second,
Linux malware is distributed among multiple architectures, as the Linux environment
itself, which also limits the number of available samples for a particular
platforma (e.g., x86 desktops). As an 
example, only 30\% of all samples 
considered in Cozzi et al's
study~\cite{cozzi2018understanding} were x86 
desktop malware samples. Finally, Linux malware is not only distributed in a self-contained form, as most Windows executables, but also as shared objects and payloads, which must be executed with a proper loader or hosting environment for its injection. 
% Finally, Linux environments evolve fast, thereby malware samples requiring old libraries and configuration files easily become outdated. 
% \chame's experiment is challenging because of these issues.

In \chame, we reverse-engineered all collected samples to 
understand their loading requirements and library dependencies. We installed all 
required library (in their outdated versions) and provided configuration
files for every sample to ensure a successful execution. From Cozzi's x86 desktop malware samples~\cite{cozzi2018understanding}, we successfully provided a suitable 
execution environment for 100 samples, and selected them for analysis. Even though this 
number is small especially compared to malware studies
in Windows, we ensured that these samples 
were diverse, fully demonstrated their malicious behaviors, and allowed the evaluation of
\chame's proof-of-concept prototype.

For future work, a usability study in \chame with a variety of benign software 
is warranted. We also plan to evaluate \chame with more diverse threats. For instance,
we plan to extend \chame protection scope to cover in-memory only 
attacks~\cite{faros} by adding support to memory-related system calls interposition
and memory snapshots acquisition, and also to detect behaviors of frequent reads and writes into the Windows export table, required by in-memory-only malware to resolve imports and exports and get the malicious payload to execute.

\textbf{\chame in Windows}:
Porting \chame Windows is necessary because Windows system is the most targeted OS by malware writers. Windows and Linux system calls are different, while there are
some correspondences between the two systems~\cite{win_lin_calls}, which are 
relevant for a future implementation of \chame in Windows. For example, for process 
control, Windows has {\tt CreateProcess}, {\tt ExitProcess} 
and {\tt WaitForSingleObject}, while Linux has {\tt fork}, 
{\tt exit} and {\tt wait}. For file manipulation, Windows 
has {\tt CreateFile}, {\tt ReadFile} and {\tt WriteFile}, 
while Linux has {\tt open}, {\tt read} and {\tt write}. 

System call hooking in Windows can be 
implemented similarly to what we
did in Linux for \chame by leveraging a 
driver that hooks the System Service 
Dispatch Table (SSDT), a task still allowed 
in 32-bit systems, but made challenging in
64-bit kernels because of the Kernel Patch
Protection (KPP) mechanism. Therefore,
\chame's implementation in 64-bit kernels 
would require Windows support. Alternatively,
\chame could be implemented in Windows 
by relying on OS callbacks and filters, 
a solution adopted by newer sandbox
solutions~\cite{Botacin2018}. For instance, a 
file-system filter provides a pre-operation
callback which allows one to deny access to a 
given file right before an access attempt, 
thus achieving the same goal of 
introducing uncertainty to OS operations. 
%SSDT is not exposed directly in Windows, but can be modified through first toggling the WP flag (default 0 for read-only) of CR0 register to gain write access of the virtual memory, and then accessing the \textit{ServiceTable} field in System Service Table (SST) pointed by \textit{KeServiceDescriptorTable} in \textit{ntoskrnl.exe}.
%The corruption protection mechanism can be implemented during the system call replacement with SSDT, through checking the parameters when a system call is invoked. The deployment of \chame on Windows will be one-time installation of the driver.}
In Windows, the {\tt EPROCESS}~\cite{eprocess}
is an equivalent structure to the Linux {\tt task\_struct}. 
Therefore, \chame can be implemented in 
Windows by adding the four fields mentioned 
earlier in Section~\ref{sec:arch} (the 
environment, the file descriptor list, the 
strategies and the threshold) to {\tt EPROCESS} to control the process environment. 
However, this modification requires access to
Windows source code and must be performed 
in-house.

Future work porting \chame to Windows OS and performing an evaluation with a larger sample size of malware is warranted.

\subsection{\chame evolves over time}

Since \chame is capable of disturbing malware execution and discovering benign software bugs, the next question becomes whether this solution could stand the test of time. We consider the decay of \chame are affected by three factors: malware detection scheme, system design and implementation, and configuration parameters.

\textbf{Malware detection scheme}: \chame targets for any hybrid detection scheme that works in a two-phase manner. Such detection scheme has been proposed more than a decade ago \cite{semantics-aware,udopayer}, and is still active in the literature \cite{symantecendpoint,li2015hybrid}. Therefore, we believe the speed for \chame to decay because of the malware detection scheme is relatively slow.

\textbf{System design}: \chame by design is a loadable kernel module. As most kernel modules or drivers, e.g. USB drivers or Bluetooth drivers, \chame requires to be updated when operating systems upgrade. For example, when {\tt sys\_call\_table} is no longer exported, a driver will have to add the {\tt kallsyms\_lookup\_name} function; when {\tt set\_memory\_rw} is no longer exported, a driver will have to implement its own. As these new functions usually require fewer than 5 lines of code, it is safe to say \chame will not decay fast because of the system design.

\textbf{Configuration parameters}: \chame uses the {\tt threshold} parameter to determine the strength of the uncertainties. As malware and benign applications evolve over time, the {\tt threshold} parameter requires to be fine tuned by the system administrator. In the future, the fine tuning can be done more efficiently with the advance of machine learning algorithms. Given the needs of organizations and companies, the machine learning algorithms can provide the most suitable set of parameters. Therefore, \chame is able to stand the test of time by integrating advanced algorithms. 

As the result from the new experiments shows, \chame is able to survive from systems ranging from Ubuntu 12.04 to Ubuntu 18.04.

\section{Related Work}\label{sec:relatedwork}
Our work intersects the areas of malware detection, software diversity and deception, and fuzz testing. This section summarizes how they have been used in software design and highlights under-studied areas.

{\bf Malware Detection Techniques} have been evolving from static,
signature-based approaches~\cite{vigna98} to dynamic,
behavior-based techniques~\cite{forrest96,semantics-aware}.
Whereas the first may be defeated by code obfuscation and malware variants,
the latter overcome these issues by continuously monitoring binaries
execution, either at API~\cite{willems2007toward,solutions2003norman} 
or system-call~\cite{bayer2006ttanalyze} levels. Dynamic solutions are 
able, for instance, to detect sensitive data leaking via system-level 
taint tracking~\cite{panorama} and keystroke logging via data-flow
analysis~\cite{martignoni08}. In this work, we leveraged the knowledge
developed by previous dynamic malware detection solutions to implement
\chame's API monitoring modules.

To detect malware, the data collected during dynamic analysis procedures 
is often modelled as behaviors and these are used as input for some
decision algorithm. Machine learning-based approaches has been leveraged 
for behavior modelling and decision with reasonable results.  Kumar et al. 
used K-means clustering~\cite{kumar2013k} to differentiate legitimate and 
malicious behaviors based on the NSL-KDD dataset. Abed et al. used bags of 
system calls to detect malicious applications in Linux containers~\cite{abed2015applying}. 
Mohaisen et al.~\cite{mohaisen2015amal} introduced AMAL to dynamically analyze 
malware using SVM, linear regression, classification trees, and kNN. 
% Fan et al. used a sequence mining algorithm to discover malicious sequential patterns and trained an All-Nearest-Neighbor (ANN) classifier based on these discovered patterns for malware detection~\cite{fan2016malicious}. 

Behaviors modelling, however, has become challenging as applications are
becoming increasingly diverse~\cite{accessminer}, which raises false
positive rates. In this scenario and as alternative for machine-learning, 
recent efforts to apply DL for malware detection have made great successes.
Pascanu et al. \cite{pascanu2015malware} used recurrent neural networks and 
echo state networks to model API system calls and C run-time library calls, 
and achieved accurate results. Li et al. leveraged an AutoEncoder and a deep 
belief network on the now outdated KDD99 dataset, and achieved a higher 
detection rate~\cite{li2015hybrid}. 
% Hou et al. constructed weighted directed graphs on collected system calls and used a deep learning framework to make dimension reduction~\cite{hou2016deep4maldroid}. 
As a drawback, current 
DL-based malware detectors work in an offline manner due to the long detection 
time and large computation resource needed. Therefore, \chame emerges as an 
alternative to bridge the gap between the efficiency of ML classifiers and the 
effectiveness of DL classifiers while monitoring binaries execution in real time. 

Most of the malware detection solutions were first implemented 
as software components, such as using patched libraries or
implementing kernel hooks, a strategy also followed by
\chame. Recently, hardware-based approaches such as 
Virtual Machine-powered solutions~\cite{dinaburg08,bromium}
emerged as alternatives for system monitoring without requiring
patching. Whereas these approaches cannot be considered
practical due to the need of developing a hypervisor, it opens
opportunity for the development of an unobtrusive \chame's implementation
in the future.

{\bf Deception:} 
% The ability to diversify behavior within a system is an essential building block for unpredictability. Diversifying components within the software stack can improve overall robustness. Researchers have studied building diverse computer systems. Forrest \etal~\cite{forrest97} proposed guidelines and advocated the use of randomized compilation techniques, which motivated later work in this area \cite{Larsen2014}. Forrest and her colleagues \cite{forrest14} also showed that code exhibits evolutionary characteristics similar to those seen in the biological world. A program, like a biological organism, has the potential to mutate but can still function normally \cite{forrest14}. 
% Several projects mitigate buffer overflows and other memory errors by randomizing system call mappings, global library entry points, stack placement, stack direction, and heap placement---often in conjunction with running multiple versions in parallel to detect divergence~\cite{chew02}. 
To a limited extent, deception has been an implicit technique for cyber warfare and defense, but is under-studied as a fundamental abstraction for secure systems. Honeypots and honeynets \cite{honeypots} are systems designed to look like production systems in order to deceive intruders into attacking the systems or networks so that the defenders can learn new techniques. 

Several technologies for providing deception have been studied. Software decoys are agents that protect objects from unauthorized access \cite{rowe}. The goal is to create a belief in the attacker's mind that the defended systems are not worth attacking or that the attack was successful. The researchers considered tactics such as responding with common system errors and inducing delays to frustrate attackers. 
Red-teaming experiments at Sandia tested the effectiveness of network deception on attackers working in groups. The deception mechanisms at the network level successfully delayed attackers for a few hours. 
Almeshekah and Spafford \cite{spafford} further investigated the adversaries' biases and proposed a model to integrate deception-based mechanisms in computer systems.  In all these cases, the fictional systems are predictable to some degree; they act as real systems given the attacker's inputs. 

True unpredictability requires randomness at a level that would cause the attacker to collect inconsistent results. This observation leads to the notion of \emph{inconsistent deception} \cite{neagoe06}, a model of deception that challenges the cornerstone of projecting false reality with internal consistency. Sun\etal~\cite{sun15spectrum,sun16bear} also argued for the value of unpredictability and deception as OS features. 
\chame explored non-intrusive unpredictable interferences to create an uncertain environment for software being deep analyzed after an initial borderline classification.

{\bf Fuzzing}: Fuzzing is an effective way to discover coding errors and security loopholes in software, operating systems, and networks by testing applications against invalid, unexpected, or random data inputs. Fuzzers can be divided into two categories: generational fuzzers, that construct inputs according to some provided format specification (e.g. SPIKE \cite{aitel2002introduction} and PEACH \cite{eddington2011peach}), and mutational fuzzers, that create inputs by randomly mutating analyst-provided or randomly-generated seeds.(e.g. AFL \cite{zalewskionline}, honggfuzz \cite{swiecki2016honggfuzz}, and zzuf \cite{hocevar2011zzuf}). Generational fuzzing requires significant manual effort to create test cases and therefore is hard to be scalable. Most of recent fuzzers are based on mutational fuzzers \cite{t-fuzz}. \chame is a mutational fuzzer that randomly applies perturbations to invoked system calls during software execution.
% and this allows different code blocks to be perturbed with one same input test case.

Fuzz testing leveraging system call behaviors has shown its potential in scalability and effectiveness.  
Trinity \cite{trinity}, for example, randomizes system call parameters to test the validation of file descriptors, and found real bugs \cite{bugsBytrinity}, including bugs in the Linux kernel. 
% \cite{2014eris, kurmuskernel, weaverperf}. 
% White-box fuzzy testers \cite{fuzzWhitebox,klee,pex} were also proposed to increase the coverage of test inputs by leveraging symbolic execution and dynamic test generation. For instance, 
BALLISTA \cite{5pointcrash} tests the data type robustness of the POSIX system call interface in a scalable way, by defining 20 data types for testing 233 system calls of the POSIX standard. \chame can also be considered as a fuzz tester at the OS system call API to understand how resilient an application is to a particular type of misbehavior. KLEE~\cite{klee} uses system call behaviors to build a model and generate high-coverage test cases to the users, and this motivated following work in coverage guided fuzzers, such as AFL \cite{zalewskionline}, honggfuzz \cite{swiecki2016honggfuzz}, and zzuf \cite{hocevar2011zzuf}, which use coverage as feedback from the target program to guide the mutational algorithm to generate inputs. While \chame's goal is not to find software bugs, \chame can borrow this idea by keeping track of a set of interesting perturbations that triggered new code paths
and focus on mutating the interesting inputs while generating new perturbations.

\section{Conclusion}\label{sec:conclusion}
This work introduces a detailed description of the design and implementation, and new extensions of \chame, a Linux framework which allows the introduction of uncertainty as an OS built-in feature to rate-limit the execution of possible malware. 
\chame allows the scheme of two-phase malware detection during software lifelong execution. After the first phase makes a borderline detection, potential malware will be disturbed when the second phase detection is under way. 

\chame offers two environments for software running in the system: (i) standard, which works according to the OS specification and (ii) uncertain, for any software that receives a borderline classification from the first phase detection, where a set of perturbations will be included.

We evaluated \chame by first manually analyzing the execution of 113 common applications and 100 malware samples from various categories. 
% We define success of software execution in the uncertain environment as benign software tolerating uncertainty and users obtaining useful results from benign software in the system. 
% Our results showed that at a threshold of 10\%, intrusive strategies thwart 62\% of malware, while non-intrusive strategies caused a failure rate of 68\%. At threshold 50\%, the percentage of adversely affected malware increased to 81\% and 76\% respectively. With a 10\% threshold, the perturbations also cause various levels of disruption (crash or hampered execution) to approximately 30\% of the analyzed benign software. With a 50\% threshold, the percentage of software adversely affected raised to 50\%. We also found that I/O-bound software was three times more affected by uncertainty than CPU-bound software. 
The results showed that a dynamic, per-system call threshold caused various levels of disruption to only 10\% of the analyzed benign software. The effects of the uncertain environment in malware was more pronounced with 92\% our studied malware samples failing to accomplish their tasks. Compared to the results obtained for a static threshold, 20\% more benign software succeeded and 24\% more malware crashed or were hampered in the uncertain environment. The results were then further corroborated by an extended dataset (5,679 Linux malware samples) on a newer system.

We also analyzed the behavior of crashed benign software, and found that many of the crashes were actually caused by software bugs. Several bugs were reproduced for Vim, tar, Mozilla Firefox and Thunderbird. 
% We provided recommendations for developers to improve software robustness, such as targeting end-to-end checks, time-consuming tests and validations. 
%Further, the uncertain environment adversely affected malware even more when the threshold increased. We also found that I/O-bound software was three times more affected by uncertainty than CPU-bound software.

% Besides effectively enabling the combination of the best of traditional machine learning and emerging deep learning methods and providing a ``safety net'' for failures of standard intrusion detection systems, \chame improves system security by (i) making systems diverse by design, (ii) increasing attackers' work factor, and (iii) decreasing the success probability and speed of attacks.
Besides filling the gap between two-phase malware detection scheme, \chame increases attackers' work factor. The effort of writing good and small malware is not lengthy endeavor. 
However, in newer system today, it is hard to use a single small malware to bypass all the protection mechanisms. The goal of \chame is to have attackers spend at least the same effort as software engineers.

The idea of making systems less predictable is audacious, nonetheless, our results indicate that an uncertain system can be feasible for raising an effective barrier against sophisticated and stealthy malware. The degree of uncertainty is not a one-size-fits-all solution---we expect an administrator to dial in the level of uncertainty to the needs of the organization and applications. 
% Finally, we define success of software execution in the uncertain environment as benign software tolerating uncertainty and users obtaining useful results from benign software in the system.

\section*{Acknowledgments} \label{sec:ack}
We would like to thank the anonymous reviewers for feedback. This work was supported through grants NSF CNS-1464801, CNS-1552059, CNS-1228839, CNS-1161541, OAC-1739025, and CNS-1747783.

\footnotesize
\bibliographystyle{plain}
\bibliography{main}

\vspace{10pt}

\section*{Biography}

\vspace{5pt}

\normalsize
\noindent\textbf{Ruimin Sun}
is a Postdoctoral Research Associate at  Northeastern University. She received Ph.D from University of Florida. She works on secure, reliable systems, and machine learning model protection.  
\vspace{5pt}

\noindent\textbf{Nikolaos Sapountzis} has a PhD from EURECOM, and currently is a Software Engineer at Cisco. His research interests include network optimization, wireless systems, privacy and security. \vspace{5pt}

\noindent\textbf{Xiaoyong Yuan} is a Ph.D. candidate at the University of Florida and specializes on machine learning and cybersecurity. 
\vspace{5pt}

\noindent\textbf{Marcus Botacin} is a Ph.D. student at Federal University
of Paran\'a. His research specializes on malware analysis and detection.
\vspace{5pt}

\noindent\textbf{Matt Bishop} is a Professor at the University of California at Davis. His research focuses on election processes, data sanitization, and the insider problem. 
%He is the author of \emph{Computer Security: Art and Science}
 \vspace{5pt}

\noindent\textbf{Donald E. Porter} is an Associate Professor at University of North Carolina at Chapel Hill. His research involves improving efficiency and security of computer systems. 
% Prior to joining UNC, he was an Assistant Professor at Stony Brook University.
\vspace{5pt}

\noindent\textbf{Andr\'e Gr\'egio} is an Assistant Professor at the Federal University of Paran\'a. His research intersects computer security and applied data science. 
\vspace{5pt}

\noindent\textbf{Xiaolin (Andy) Li} is a Partner of Tongdun Technology, heading the AI Institute. His research interests include deep learning, cloud computing, and security \& privacy.
\vspace{5pt}

\noindent\textbf{Daniela Oliveira} is an Associate Professor at University of
Florida. Her research includes human factors security and IoT security.
\vspace{5pt}

\end{document}